\documentclass[aps,prd,superscriptaddress,nofootinbib,tighten,preprint]{revtex4}
\usepackage[utf8]{inputenc}
\usepackage[english]{babel}
\usepackage{amsmath}
\usepackage{subfigure}
\usepackage{amsfonts}
\usepackage{amssymb}
\usepackage{graphicx}
\newcommand{\be}{\begin{equation}}
\newcommand{\ee}{\end{equation}}
\newcommand{\bea}{\begin{eqnarray}}
\newcommand{\eea}{\end{eqnarray}}

\newcommand{\nn}{\nonumber}

\catcode`@=12

\begin{document}

\title{Axion and FIMP Dark Matter in a $U(1)$ extension of the Standard Model}

\author{Laura Covi} 
\author{Sarif Khan}

\affiliation{Institute for theoretical physics, Georg-August University G\"ottingen, D-37077 Germany}

\begin{abstract} 
In the Standard Model a Dark Matter candidate is missing, but it is relatively 
simple to enlarge the model including one or more suitable particles. 
We consider in this paper one such extension, inspired by simplicity and 
by the goal to solve more than just the Dark Matter issue. 
Indeed we consider a local $U(1) $ extension of the SM providing an
axion particle to solve the strong CP problem and including RH neutrinos 
with appropriate mass terms. One of the latter is decoupled from the SM 
leptons and can constitute stable sterile neutrino DM.
In this setting, the PQ symmetry arises only as an accidental symmetry
but its breaking by higher order operators is sufficiently suppressed to
avoid introducing a large $ \theta $ contribution.
The axion decay constant and the RH neutrino masses are related
to the same v.e.v.s and the PQ scale and both DM densities are determined by
the parameters of the axion and scalar sector.  
The model predicts in general a mixed Dark Matter 
scenario with both axion and sterile neutrino DM and is characterised by 
a reduced density and observational signals from each single component.
\end{abstract}
\maketitle

\section{Introduction}

Till date, the Standard Model (SM) has been a very successful theory in describing 
nature and with the recent discovery of the Higgs boson it is now a complete model
that can be extended to higher scale \cite{Buttazzo:2013uya}.
Nevertheless, the SM cannot be the full theory as it does not address a few 
important issues, e.g. the absence of a Dark Matter candidate, whose presence
is seen in many astrophysical and cosmological observations \cite{Bertone:2016nfn},
or the generation of neutrino masses and mixings~\cite{Kajita:2016cak, McDonald:2016ixn, Esteban:2020cvm}.
In this paper we add to the Standard Model a new dark sector in order to give
a concerted solution to these issues as well as to the strong CP problem,
i.e. the very long standing puzzle connected with the suppression of the following term in the SM lagrangian
\begin{eqnarray}
\mathcal{L}^{\theta}_{CP} &=& \left(\theta + {\rm arg[det(m_{q})]}\right) \frac{g^2_{s}}{32 \pi^2} G_{\mu\nu} \tilde{G^{\mu\nu}} \nn \\
&=& \bar{\theta} \frac{g^2_{s}}{32 \pi^2} G_{\mu\nu} \tilde{G^{\mu\nu}}
\label{theta-g-g}
\end{eqnarray}   
where $\tilde{G^{\mu\nu}} = \epsilon^{\mu\nu\alpha\beta} G_{\alpha\beta}$ and $m_{q}$ is
the mass matrix for the SM quarks. 
This term breaks explicitly the Parity symmetry and generates a non-vanishing
electric dipole moment for the neutron of the order $ d_n \sim 10^{-16}\; \bar\theta \, {\rm e\,cm} $.
The present experimental bound $ d_n < 10^{-26}\,\, {\rm e\,cm}$,
puts a very strong bound on the $\theta-$term {\it i.e} $\bar \theta < 10^{-10}$ \cite{Baker:2006ts}.
Such small value for the parameter $ \bar \theta $ is unnatural, as $ \bar \theta $ is an
angular variable, including both the vacuum and the EW contributions, and is expected to be of order 
$ {\cal O} (1) $ \footnote{Note that if we arrange to have $\bar \theta \sim 0$ at a high scale, 
then in the SM loop corrections can only generate  $\bar{\theta} < 10^{-17}$ \cite{ellis, georgi}. 
This of course is different in SM extensions like supersymmetry where additional CP violating
phases and fields are present~\cite{Arbey:2019pdb}.}. 

One of the most famous and still viable solutions of this problem is the introduction of an axial global 
symmetry namely the Peccei-Quinn symmetry~\cite{Peccei:1977hh, Peccei:1977ur, Weinberg:1977ma, Wilczek:1977pj}
broken spontaneously at a high scale $ F_a$. In that context the $\theta$-term becomes
dynamical and it relaxes to the minimum of potential (at $\bar \theta \sim 0$) 
in the course of the cosmological evolution. 
In order to achieve a dynamical $ \theta $ we need a new pseudoscalar field and therefore we have to extend
the SM to contain an additional extra CP-odd degree of freedom, the axion field. But as global symmetries
are not respected by gravity \cite{Hawking:1987mz, Lavrelashvili:1987jg, Giddings:1988cx, Coleman:1988tj, Gilbert:1989nq, Banks:2010zn}, 
we expect such a PQ symmetry to be only accidental and to be broken by non-renormalisable operators 
at the Planck scale.

It is well-known that the axion can be a good Dark Matter candidate \cite{Preskill:1982cy, Abbott:1982af}, 
solving another of the open issues of the SM, and that in this case $ F_a \sim 10^{10-11} $ GeV has to be 
at an intermediate scale between the EW scale and the Planck scale. Such a mass scale
can also fit with the mass of the RH neutrinos and so the light neutrino masses could 
be sourced by the seesaw mechanism to obtain active neutrino masses in the sub-eV scale.
The neutrino masses have been addressed in the context of axion models in 
\cite{Clarke:2015bea, Salvio:2015cja}.   

In the present work, we propose a new realisation of a gauged $ U(1) $ extension
of the SM which contains an accidental PQ symmetry and an axion field.
We add to the SM gauge group an additional local gauge group $U(1)_{X}$ and 
a discrete symmetry $\mathbb{Z}_{2}$. The gauge symmetry is broken at a
high scale by two complex scalar v.e.v.s such that one pseudoscalar field
remains physical and plays the role of the axion.
Moreover, apart for the new gauge boson and the new scalar sector, we 
enlarge the particle spectrum by three RH neutrinos and two sets of Dirac fermions,
the latter charged under the color gauge group as well as the $ U(1)_X $.
We investigate the anomaly constraints of the $ U(1)_X$ and we solve
those conditions fixing the new charges such that the operators providing the  
RH neutrino masses are present in the lower orders in the same v.e.v.s. 

The charges of the $ U(1)_X $ field are also such that a global PQ
symmetry will accidentally emerge from this local $ U(1)_X$. Moreover
the global $U(1) $ symmetry has a color anomaly due to the
presence of the additional colored fermions and generates the $ \theta-$term
for the axion field after the exotic quarks are integrated out.
After the QCD phase transition a potential for the axion arises from 
QCD instantons~\cite{DiLuzio:2020wdo}.
Due to gravitational effects not respecting the global symmetry, 
we expect higher dimensional operators at the Planck scale which 
violate the global PQ-symmetry but conserve the gauged ($U(1)_X$) symmetry.
Because of this extra contribution from higher dimensional operator, the 
axion minimum shifts from the value $\bar{\theta} = 0$.  Such shift 
is restricted by the bound on the neutron electric dipole moment and 
so  also the range of the scalar v.e.v.s will be bounded from above.
As we will see, reaching the value of $ F_a $ compatible with the full
DM density via the misalignment mechanism \cite{Preskill:1982cy, Abbott:1982af} 
is not always possible in the model, but luckily,  if we assume the RH neutrinos 
mass to be at the  GeV scale and the $U(1)_X$ gauge coupling very small, 
we can consider one of the RH neutrinos as a feebly interacting massive 
particle (FIMP) DM component. 
We will explore in the following the full parameter space of the model
and see that in most regions a mixed DM density arises, lowering
possible axion detection signals as the FIMP fills the gap to the
observed density.

The paper is organised in the following way. 
In Section \ref{model:anomaly}, we have addressed the
model in detail including the discussion of the anomaly constraints,
in Section \ref{neutrino} we show how we can fit the present
neutrino data within the model with only two RH neutrinos.
In Section \ref{axion-part} we have a look on the axion field in the
model and compute its relic density by misalignment.
In Section \ref{FIMP} we consider the lightest RH neutrino
as a FIMP DM candidate and show that it can fill the gap
to reach the observed DM density. Finally we conclude in Section \ref{conclusion}.

\section{The Model}
\label{model:anomaly}

In the present work, we introduce a $U(1)_{X}$ gauge extension of the SM, together with 
an extended particle content with three right-handed neutrinos $N_{1}$ and $N_{2}, N_{3} $ 
with $U(1)_{X}$ charges $n_e$ and $n$, two sets of exotic quarks
$\psi_L^i, \psi_R^i$, and $\chi_{L}^i, \chi_{R}^i $ with charges under $U(1)_X$ as 
$\alpha_L$, $\alpha_R$, $\beta_L$ and $\beta_R$, respectively, as well as
two singlet scalars $\phi_1$, $\phi_2$ 
with $U(1)_{X}$ charges $\alpha_L - \alpha_R$, $\beta_L - \beta_R$.
Here we consider different charges in the RH neutrino sector such that one of the RH neutrinos
does not couple to the light leptons and remains as a possible DM candidate.
Moreover the charges of the scalars are fixed to allow a Yukawa coupling with the
two sets of fermions generating the fermion masses at the PQ scale $ F_a$.
The Standard Model leptons have the same charge as the corresponding RH neutrino,
completing a vectorial representation of the $U(1)_X$ symmetry and similarly 
the SM quarks all have the same vectorial charge $m$.

Among the additional particles, only the two sets of exotic quarks transform non trivially 
under $SU(3)_c$  gauge group, while all the rest are SM singlets. 
We introduce as well a  $\mathbb{Z}_{2}$
symmetry  which ensures  the lightest RH neutrino $N_1$ as stable dark matter candidate 
and forbids the mixing among the extra fermions with the same QCD charge. 
In Table\,(\ref{tab1}, \ref{tab2}), 
we show all the SM and beyond SM particles with their corresponding charges under the complete 
gauge group $SU(3)_{c} \times SU(2)_{L} \times U(1)_{Y} \times U(1)_{X} \times \mathbb{Z}_2$.   
\begin{center}
\begin{table}[h!]
\begin{tabular}{||c|c|c|c||}
\hline
\hline
\begin{tabular}{c}
    Gauge\\
    Group\\ 
    \hline
    
    ${\rm SU(2)}_{\rm L}$\\ 
    \hline
    ${\rm U(1)}_{\rm Y}$\\ 
    \hline
    $U(1)_{X}$\\ 
    \hline
    $U(1)_{PQ}$\\ 
\end{tabular}
&

\begin{tabular}{c|c|c}
    \multicolumn{3}{c}{Baryon Fields}\\ 
    \hline
    $Q_{L}^{i}$&$u_{R}^{i}$&$d_{R}^{i}$\\ 
    \hline
    $2$&$1$&$1$\\ 
    \hline
    $1/6$&$2/3$&$-1/3$\\ 
    \hline
    $m$&$m$&$m$\\
    \hline
    $0$&$0$&$0$\\     
\end{tabular}
&
\begin{tabular}{c|c|c|c|c|c|}
    \multicolumn{6}{c}{Lepton Fields}\\
    \hline
   $L_{L}^{e}$ & $L_{L}^{\mu}$ &$L_{L}^{\tau}$&$e_{R}$&$\mu_{R}$&$\tau_{R}$\\
    \hline
    $2$&$2$&$2$&$1$&$1$&$1$\\
    \hline
    $-1/2$&$-1/2$&$-1/2$&$-1$&$-1$&$-1$\\
    \hline
    $n_e$&$n$&$n$&$n_e$&$n$&$n$\\
        \hline
    $- 2 q_a$&$0$&$0$&$-2 q_a$&$0$&$0$\\
\end{tabular}
&
\begin{tabular}{c}
    \multicolumn{1}{c}{Scalar Fields}\\
    \hline
    $\phi_{h}$\\
    \hline
    $2$\\
    \hline
    $1/2$\\
    \hline
    $0$\\
    \hline
    $0$\\
\end{tabular}\\
\hline
\hline
\end{tabular}
\caption{SM particles and their corresponding charges under complete gauge group. All the particles are even under
$\mathbb{Z}_2$ discrete gauge group. The doublets are
defined as $Q_{L}^{i}=(u_{L}^{i},d_{L}^{i})^{T}$ and $Q_{L}^{i}=(u_{L}^{i},d_{L}^{i})^{T}$, respectively.}
\label{tab1}
\end{table}
\end{center}

\begin{center}
\begin{table}[h!]
\begin{tabular}{||c|c|c||}
\hline
\hline
\begin{tabular}{c}
    Gauge\\
    Group\\ 
    \hline
    
    ${\rm SU(3)_{c}, SU(2)}_{\rm L}$\\ 
    \hline
    $U(1)_{X}$\\ 
    \hline
    $U(1)_{PQ}$\\
    \hline
    $\mathbb{Z}_2$\\ 
    \hline
    No. of flavors\\ 
\end{tabular}
&

\begin{tabular}{c|c|c|c|c|c|c}
    \multicolumn{7}{c}{Fermions}\\ 
    \hline
    $N_1$&$N_2$&$N_3$ & $\psi_L$& $\psi_R$& $\chi_L$&$\chi_{R}$\\ 
    \hline
    $(1,1)$&$(1,1)$&$(1,1)$&$(3,1)$&$(3,1)$&$(3,1)$&$(3,1)$\\ 
    \hline
    $n_e$&$n$&$n$&$\alpha_L$&$\alpha_R$&$\beta_L$&$\beta_R$\\
    \hline
    $-2 q_a$&$0$&$0$&$-q_a$&$q_a$&$q_a$&$-q_a$\\
    \hline
    $-1$&$1$&$1$&$1$&$1$&$-1$&$-1$\\
    \hline
    $1$&$1$&$1$&$N_{\psi}$&$N_{\psi}$&$N_{\chi}$&$N_{\chi}$\\ 
\end{tabular}
&
\begin{tabular}{c|c}
    \multicolumn{2}{c}{Scalars}\\
    \hline
     $\phi_1$ & $\phi_2$\\
    \hline
    $1$&$1$\\
    \hline
    $\alpha_L - \alpha_R$ & $\beta_L - \beta_R$\\
    \hline
     $- 2 q_a$ & $2 q_a$\\
     \hline
     $1$ & $1$\\
     \hline
     $1$ & $1$\\
\end{tabular}\\
\hline
\hline
\end{tabular}
\caption{BSM particles and their corresponding charges under complete gauge group and global $U(1)_{PQ}$ symmetry.
The $U(1)_Y $ charges are vanishing and therefore not given in the table.}
\label{tab2}
\end{table}
\end{center}

\subsection{Gauge Anomaly Cancellation}

In order to fix the $U(1)_X $ charges of the new particles and the SM particles, we need to 
consider the gauge  anomaly involving the new local gauge group and the SM gauge group, 
i.e. the following gauge  combinations:
$\left[SU(3)\right]^{2} \times U(1)_{X}$, 
$\left[SU(2)\right]^{2} \times U(1)_{X}$,
$\left[ U(1)_{Y} \right]^{2} \times U(1)_{X}$, $\left[ U(1)_{X}\right]^{3}$,
$\left[ U(1)_{X} \right]^{2} \times U(1)_{Y}$ and 
$\left[Gravity \right]^{2} \times U(1)_{X}$\,.
As we have chosen the RH neutrinos and the SM leptons to have vectorial
charges under the new symmetry, they do not contribute to most of the anomaly
constraints.
The following non-trivial conditions on the $U(1)_{X}$ charges arise: 
\begin{itemize}
\item from $\left[Gravity \right]^{2} \times U(1)_{X}$ we obtain
\begin{eqnarray}
 3 N_{\psi} (\alpha_{L} - \alpha_{R}) &=& 
 3 N_{\chi} (\beta_{R} - \beta_{L}) \nonumber \\
(\alpha_{L} - \alpha_{R}) &=& 
n_{\chi} (\beta_{R} - \beta_{L})
\label{gravity-u1x}
\end{eqnarray}
where $n_{\chi} = \frac{N_{\chi}}{N_{\psi}}$.

\item from $\left[SU(3) \right]^{2} \times U(1)_{X}$ we obtain
\begin{eqnarray}
 N_{\psi} (\alpha_{L} - \alpha_{R}) &=& 
 N_{\chi} (\beta_{R} - \beta_{L}) \nonumber \\
(\alpha_{L} - \alpha_{R}) &=& 
n_{\chi} (\beta_{R} - \beta_{L})
\label{su3-u1x}
\end{eqnarray}
same as the previous constraint.

\item the $\left[U(1)_{X} \right]^{3}$ anomaly gives
\begin{eqnarray}
 N_{\psi} (\alpha_{L}^3 - \alpha_{R}^3) &=& 
 N_{\chi} (\beta_{R}^3 - \beta_{L}^3) \nonumber \\
 \alpha^2_{L} + \alpha_{L} \alpha_{R} + \alpha^{2}_{R} &=& \beta^2_{L}
+ \beta_{L} \beta_{R} + \beta^2_{R}\,, 
\label{u1x3}
\end{eqnarray} 
where in the last line we have used Eq.\,(\ref{gravity-u1x}) to simplify the equation to a
second order equation in the charges.

\item for the combinations $\left[SU(2) \right]^{2} \times U(1)_{X}$ and
 $\left[U(1)_{Y}  \right]^{2} \times U(1)_{X}$ only the SM states contribute and they give 
 \begin{eqnarray}
 9\; m + n_{e} + 2 n = 0\,.
 \label{su2u1x}
 \end{eqnarray}
\end{itemize}

Now let us consider more in detail the constraints on the exotic quark charges.
We can rewrite  Eq.\,(\ref{u1x3}) to depend only on the charge difference
$\Delta \beta = \beta_{R} - \beta_{L}$ and the charge ratios $ y =  \frac{\Delta \beta}{\alpha_{R}},
 z = \frac{\beta_L}{\alpha_{R}}$  as
\begin{eqnarray}
&&(n^2_{\chi} - 1) (\beta_{R} - \beta_{L})^{2} + 3 (\beta_{R} - \beta_L)
(\alpha_{R} n_{\chi} - \beta_{L}) + 3 (\alpha^2_{R} - \beta^2_{L}) = 0\, , \nonumber \\
&& \rightarrow (n^2_{\chi} - 1) y^{2} + 3 y (n_{\chi} -z )  + 3 (1 - z^{2}) = 0\,.
\label{quadratic-equation}
\end{eqnarray}
So here we have a quadratic equation in both $y$ and $z$ as long as $ n_\chi \neq 1 $. 
The special case $ N_\psi = N_\chi $ is not interesting since then we have opposite charges 
for the new scalar fields and then the PQ symmetry coincides with the $ U(1)_X $. 


The solutions of the above quadratic equation for $y$ are given as,
\begin{eqnarray}
y_{\pm} = \frac{- 3 (n_{\chi} - z) \pm \sqrt{9 (n_{\chi} - z)^{2} - 
12 (n^2_{\chi} -1) (1 - z^{2}) }}{2 (n^2_{\chi} -1)}\, ,
\label{y-eigenvalues}
\end{eqnarray}
and they simplify to a single non-vanishing rational root for $ z = \pm 1 $, i.e.
\begin{eqnarray}
y_{-} =  - \frac{3}{n_{\chi} \pm 1}\, .
\label{y-singleroot}
\end{eqnarray}
The other solution $ y_{+} = 0 $ just corresponds to vectorial charges for the exotic 
fermions and therefore vanishing $X$ charge for the exotic scalar fields.

 A physical solution has to be real, so the discriminant must be positive,
 giving the bound
 \begin{eqnarray}
 (4 n^{2}_{\chi} - 1) z^{2} - 6 n_{\chi} z - n^{2}_{\chi} + 4 \geq 0\,.
 \label{z-equation}
 \end{eqnarray}
 The above equation is quadratic in $z$ and has always real solutions,
 given by
  \begin{eqnarray}
 z_{\pm} = \frac{3 n_{\chi} \pm 2 |n^2_{\chi} - 1|}{(4 n^2_{\chi} - 1)}
 \end{eqnarray}
 Depending on the sign of $4 n^{2}_{\chi} - 1$, different ranges for $ z $
are allowed , i.e.
 \begin{itemize}
 \item when $n_{\chi} > 1/2 $, we must require $ z \leq z_- $ or $ z \geq z_+ $;
as for large $ n_{\chi} $, $ z_{\pm} \rightarrow \pm 1/2 $ this reduces to
the requirement $ |z| > 1/2 $ in the limit $n_{\chi} \gg 1$\,,
 \item when $ n_{\chi} < \frac{1}{2}$, we must require
 $ z_- \leq z \leq z_+ $ and this reduces to $|z| > 2$ for
 $n_{\chi} \ll 1$\,. 
 \end{itemize}
 Without loss of generality, we assume that $n_{\chi} > 1$
because the case $n_{\chi} < 1$  is the same just exchanging the role
of the two sets of exotic fermions, i.e. $ \alpha_i \leftrightarrow \beta_i ,
N_\psi \leftrightarrow N_\chi, n_\chi \leftrightarrow 1/n_\chi $. 

 We can now determine the different values of $\alpha_{L,R}$
 and $\beta_{L,R}$ fixing the relative number of exotic quarks
 $n_{\chi}$ and the ratio $z$. For the case $ |z| = 1 $, the non trivial solutions 
 are unique and rational and we concentrate on this particular case. 
 We show in Table \ref{u1xcharge} the $U(1)_X$ the solutions of the
 anomaly conditions for different values of $n_{\chi}$, $z = \pm 1$ and 
 one can easily determine the charges for other cases as well
 as a function of one reference charge $ \alpha_R $ 
 using eqs.~(\ref{su3-u1x}), (\ref{y-singleroot}) and the definitions of $ y, z$.
 Note that the value of $ n_\chi $ has to be large in order to obtain very
 different charges for the scalar fields and so suppress all the possible
 higher dimensional operators in the scalar potential.
 
Now we are left to fix the charges of the SM fields.
In order to be able to write down the Dirac and Majorana mass terms associated
with the left and right handed neutrinos as couplings with the
exotic scalars, we fix $n_{e}$ and $n$ in terms of the charges of the exotic quarks 
as $n_{e} = \frac{1 + n_{\chi}}{2} (\beta_{R} - \beta_{L}) = \frac{1 + n_{\chi}}{2}y\; \alpha_R $ 
and 
$n =  \frac{1- n_{\chi}}{2} (\beta_{R} - \beta_{L}) =   \frac{1 - n_{\chi} }{2} y\; \alpha_R $.
Finally from eq.~(\ref{su2u1x}), the quark charge has to be 
$m = - \frac{n_e+ 2 n}{9} =  \frac{n_\chi - 3}{18}  y \; \alpha_R  $.

 \begin{center}
\begin{table}[h!]
\begin{tabular}{||c|c|c|c|c|c|c|c|c|c|c||}
\hline
\hline
\begin{tabular}{c}
    $n_{\chi}$\\ 
    \hline
     10\\ 
    \hline
    10\\ 
    \hline
    11\\ 
    \hline
    11\\
\end{tabular}
&

\begin{tabular}{c}
    $z$\\ 
    \hline
    
    $1$\\ 
    \hline
    $-1$\\ 
    \hline
    $1$\\ 
    \hline
    $-1$\\
\end{tabular}
&
\begin{tabular}{c}
$y$\\
    \hline
    
    $-\frac{3}{11}$\\ 
    \hline
    $-\frac{1}{3}$\\ 
    \hline
    $-\frac{1}{4}$\\ 
    \hline
    $-\frac{3}{10}$\\
\end{tabular}
&

\begin{tabular}{c}
    $\alpha_{L}$\\ 
    \hline
    
    $-\frac{19}{11} \alpha_{R}$\\ 
    \hline
    $-\frac{7}{3} \alpha_R$\\ 
    \hline
    $-\frac{7}{4} \alpha_R$\\ 
    \hline
    $-\frac{23}{10} \alpha_R$\\
\end{tabular}
&

\begin{tabular}{c}
    $\beta_L$\\ 
    \hline
    
    $\alpha_R$\\ 
    \hline
    $-\alpha_R$\\ 
    \hline
    $\alpha_R$\\ 
    \hline
    $-\alpha_R$\\
\end{tabular}
&

\begin{tabular}{c}
    $\beta_R$\\ 
    \hline
    
    $\frac{8}{11} \alpha_R$\\ 
    \hline
    $- \frac{4}{3} \alpha_R$\\ 
    \hline
    $\frac{3}{4} \alpha_R$\\ 
    \hline
    $-\frac{13}{10} \alpha_R$\\
\end{tabular}
&

\begin{tabular}{c}
    $\alpha_L- \alpha_R$\\ 
    \hline
    
    $ - \frac{30}{11} \alpha_R$\\ 
    \hline
    $- \frac{10}{3} \alpha_R$\\ 
    \hline
    $ - \frac{11}{4} \alpha_R$\\ 
    \hline
    $-\frac{33}{10} \alpha_R$\\
\end{tabular}
&

\begin{tabular}{c}
    $\beta_L- \beta_R$\\ 
    \hline
    
    $\frac{3}{11} \alpha_R$\\ 
    \hline
    $ \frac{1}{3} \alpha_R$\\ 
    \hline
    $\frac{1}{4} \alpha_R$\\ 
    \hline
    $ \frac{3}{10} \alpha_R$\\
\end{tabular}
&

\begin{tabular}{c}
    $n_e$\\ 
    \hline
    
    $- \frac{3}{2} \alpha_R$\\ 
    \hline
    $- \frac{11}{6} \alpha_R$\\ 
    \hline
    $- \frac{3}{2} \alpha_R$\\ 
    \hline
    $-\frac{9}{5} \alpha_R$\\
\end{tabular}
&

\begin{tabular}{c}
    $n $\\ 
    \hline
    
    $\frac{27}{22} \alpha_R$\\ 
    \hline
    $ \frac{3}{2} \alpha_R$\\ 
    \hline
    $\frac{5}{4} \alpha_R$\\ 
    \hline
    $ \frac{3}{2} \alpha_R$\\
\end{tabular}
&

\begin{tabular}{c}
    $m$\\ 
    \hline
        $-\frac{7}{66} \alpha_R$\\ 
    \hline
    $- \frac{7}{54} \alpha_R$\\ 
    \hline
    $ -\frac{1}{9} \alpha_R$\\ 
    \hline
    $-\frac{2}{15} \alpha_R$\\
\end{tabular}
\\
\hline
\hline
\end{tabular}
\caption{$U(1)_{X}$ charges of the exotic quarks and SM fields 
for different values of $n_{\chi}$ and $z$ in terms of the charge $\alpha_R$.}
\label{u1xcharge}
\end{table}
\end{center}

So far the total number of exotic quarks $ N_\psi + N_\chi = N_\psi (1+ n_\chi ) $ has not been
fixed and the $U(1)_X$ charges only depend on the relative number $ n_\chi $, but we will see 
later that in order to suppress higher order operators in the axion potential this number has to be
large. Of course the new exotic quarks will affect the running of the $ SU(3) $ coupling above
their mass scale and contribute to the beta function as the SM quarks, i.e. at one loop we have
\begin{equation}
\beta_3 (\alpha_3) = - \frac{\alpha_3^2}{2\pi} \left[ 7 - \frac{2 (N_\psi + N_\chi)}{3}  \right]
\end{equation} 
so that to keep asymptotic freedom only up to 10 exotic quarks are allowed. 
We will discuss later if this constraint can be fulfilled.

With the charge assignments in Table \ref{u1xcharge}, the complete Lagrangian for the particles in the 
Table \ref{tab1} and Table \ref{tab2} is given by
\begin{eqnarray}
\mathcal{L} &=& \mathcal{L}_{SM} + \sum_{j = 1,2,3} 
\frac{i}{2} \bar{N_{j}} \gamma^{\mu} D_{\mu} N_{j} 
+ \sum^{N_{\psi}}_{j=1} i \bar{\psi_j} \gamma^{\mu} D_{\mu} \psi_j
+ \sum^{N_{\chi}}_{j=1} i  \bar{\chi^{j}} \gamma^{\mu} D_{\mu} \chi^{j} 
+ \sum^{2}_{j = 1}(D_{\mu} \phi_j)^{\dagger} (D^{\mu} \phi_j)  \nn \\
&& + \mathcal{L}^{Yuk}_{BSM} + \mathcal{L}^{Yuk}_{N} + \mathcal{V}(\phi_{h},\phi_i )
\label{lagrangian-model}
\end{eqnarray}
where $D_{\mu}$ corresponds to the co-variant derivative and depends on the different charge assignments of
the fields. $\mathcal{L}_{SM}$ represents the Lagrangian for the SM particles, the next terms 
are the kinetic terms for right handed neutrinos, extra fermions and singlet scalars, respectively. 
$\mathcal{L}^{Yuk}_{N}$ contains the Yukawa couplings for leptons and right handed neutrinos including
operators to dimension 5 suppressed by the Planck scale.  These are as follows
\begin{eqnarray}
\mathcal{L}^{Yuk}_{N} &=&  \sum_{i = 2,3}  \left[ y_{\mu i} \bar L_{\mu} \phi_{h} N_{i}
 + y_{\tau i} \bar L_{\tau} \phi_{h} N_{i}
+ y_{e i} \bar L_{e} \phi_{h} N_{i} 
\frac{\phi_{1}}{M_{PL}} \right]
\nn \\
 & &
+  \sum_{i,j = 2,3} \frac{y_{ij} }{2} N_{i} N_{j}  \frac{\phi_{1} \phi_{2}}{M_{PL}}
+ \frac{y_{11}}{2}  N_{1} N_{1} 
\frac{\phi_{1}^\dagger  \phi_{2}}{M_{PL}} + {\it h.c.} .
\label{hdo-rh-neutrino}
\end{eqnarray}  
We see here clearly that the $ N_1 $ RH neutrino is singled out by having a different $ U(1)_X $ and $ \mathbb{Z}_2  $ 
charges and cannot mix with the other two, nor with the light neutrinos.
The last terms are higher dimensional operators for the right handed neutrinos which generate a mass once 
$\phi_{1,2}$ develop vevs and will be discussed in the next section.   

The term $\mathcal{L}^{Yuk}_{BSM}$ represents the Yukawa sector for the additional exotic quarks and takes 
the following form
\begin{eqnarray}
\mathcal{L}^{Yuk}_{BSM} &=& \sum^{N_{\psi}}_{i, j = 1} \lambda_{ij} \, \bar{\psi}^i_{L} \psi^j_{R} \phi_{1}
+ \sum^{N_{\chi}}_{i, j = 1} y_{ij} \bar{\chi}^{i}_{L} \chi^{j}_{R} \phi_2  + {\it h.c.}\,.
\end{eqnarray}

Here also we can forbid the cross terms between the two sectors assuming opposite $\mathbb{Z}_2  $ 
parity~\footnote{As these states are heavy and can be integrated out, 
any mixing term with the light quarks will generate new sources of flavour violation via higher dimensional 
operators, suppressed by the high masse scale  $ \sim F_a $, and therefore possibly not too dangerous, 
but the charges can be chosen to avoid them at least at level of dimension six operators.}.
These Yukawa terms are important because they are compatible with an accidental global Peccei-Quinn
symmetry and by assigning a Peccei-Quinn charge to the exotic fermions and scalars, we can identify
one of the pseudoscalars with the axion and generate the dimension five operator coupling of such an 
axion with the gluons.  So our model is a generalisation of the KSVZ type of axion models with multiple 
exotic quarks~\cite{Kim:1979if, Shifman:1979if}.

In  order to break the PQ symmetry as well as the $ U(1)_X $ symmetry, the scalar fields
have to take a vacuum expectation value.  The complete potential
$\mathcal{V} (\phi_{h}, \phi_i)$
up to mass dimension four has the following form,
\begin{eqnarray}
\mathcal{V}(\phi_{h},\phi_i ) &=& -\mu^2_{h} (\phi^{\dagger}_h \phi_h)
+ \lambda_h (\phi^{\dagger}_h \phi_h)^2 + \sum_{i = 1,2} \left[-\mu^2_{\phi_i} (\phi^{\dagger}_i \phi_i)
+ \lambda_{\phi_i} (\phi^{\dagger}_i \phi_i)^2 \right]
 \nn \\ 
 &+& \sum_{i = 1,2} \lambda_{h \phi_i} (\phi^{\dagger}_h \phi_h) (\phi^{\dagger}_i \phi_i)
 + \lambda_{12} (\phi^{\dagger}_1 \phi_1) (\phi^{\dagger}_2 \phi_2)\,.
\end{eqnarray}
In principle also higher dimensional operators could appear here, but mostly they depend just on 
$ | \phi_i |^2 $ due to the different charge assignments. 
These terms would just change slightly the vacuum expectation values $ v_i $, but remain invariant
under the phase shift of the scalar fields and therefore do not depend on the axion.
Other higher dimensional operators affecting the axion have to be suppressed in order not to
spoil the solution of the strong CP problem and they will be discussed in Section \ref{axion-part}.

After spontaneous symmetry breaking, the scalars can be written 
as
\begin{eqnarray}
\phi_{h}=
\begin{pmatrix}
0 \\
\dfrac{v+H}{\sqrt{2}}
\end{pmatrix}
\,,\,\,
\phi_{1}=
\begin{pmatrix}
\dfrac{v_1 + H_{1} + i a_1}{\sqrt{2}}
\end{pmatrix}\,,
\,\,
\phi_{2}=
\begin{pmatrix}
\dfrac{v_2 + H_{2} + i a_2}{\sqrt{2}}
\end{pmatrix}\,.\,\,
\label{phih}
\end{eqnarray}
We postpone the discussion of the CP-odd component of the singlet scalars till we reach Section \ref{axion-part}.
The tad-pole conditions for the potential mentioned above are as follows
\begin{eqnarray}
\mu^2_h &=& \lambda_h v^2 + \lambda_{h \Phi_1} \frac{v^2_1}{2} + \lambda_{h \Phi_2} \frac{v^2_2}{2} \nn \\
\mu^2_{\Phi_1} &=& \lambda_{\Phi_1} v^2_1 + \lambda_{h \Phi_1} \frac{v^2}{2} + \lambda_{12} \frac{v^2_2}{2} \nn \\
\mu^2_{\Phi_2} &=& \lambda_{\Phi_2} v^2_2 + \lambda_{h \Phi_2} \frac{v^2}{2} + \lambda_{12} \frac{v^2_1}{2}
\end{eqnarray} 

\subsection{Neutral Scalars}

Once the $U(1)_X $, the PQ and electro-weak symmetry break completely, using the above tadpole conditions,
we can write down the mass matrix for the neutral scalars
in the basis ($H, H_1, H_2$) in the following way,
\begin{eqnarray}
M^2_{S} =
\begin{pmatrix}
 2 \lambda_h v^2 & \lambda_{h \Phi_1} v v_1 & \lambda_{h \Phi_2} v v_2  \\
\lambda_{h \Phi_1} v v_1 & 2 \lambda_{\Phi_1} v^2_1 & \lambda_{12} v_1 v_2 \\
\lambda_{h \Phi_2} v v_2 & \lambda_{12} v_1 v_2 & 2 \lambda_{\Phi_2} v^2_2
\end{pmatrix}\,.
\label{neural-mass-higgs}
\end{eqnarray}
From Eq.\,(\ref{neural-mass-higgs}), one can see that to get the mass eigenstates for the neutral scalars ($h, h_1, h_2$),
then one needs to diagonalize the above mass matrix. As we know by the Euler angles
($\theta^{\prime}_{01}, \theta^{\prime}_{02}, \theta^{\prime}_{12}$) we can make the $3 \times 3$ matrix in the diagonal form
and the two basis are related in the following way,
\begin{eqnarray}
\begin{pmatrix}
h \\
h_1 \\
h_2
\end{pmatrix} 
=
\begin{pmatrix}
c^{\prime}_{01} c^{\prime}_{02} & s^{\prime}_{01} c^{\prime}_{02} & s^{\prime}_{02} \\
-s^{\prime}_{01} c^{\prime}_{12} - c^{\prime}_{01} s^{\prime}_{12} s^{\prime}_{02} & c^{\prime}_{01} c^{\prime}_{12} - s^{\prime}_{01} s^{\prime}_{12} s^{\prime}_{02} & s^{\prime}_{12} c^{\prime}_{02} \\
s^{\prime}_{01} s^{\prime}_{12} - c^{\prime}_{01} c^{\prime}_{12} s^{\prime}_{02} & -c^{\prime}_{01} s^{\prime}_{12} - s^{\prime}_{01} c^{\prime}_{12} s^{\prime}_{02} & c^{\prime}_{12} c^{\prime}_{02}
\end{pmatrix}
\begin{pmatrix}
H \\
H_1 \\
H_2
\end{pmatrix}
\end{eqnarray}
where $c^{\prime}_{ij} = \cos \theta^{\prime}_{ij}$ and $s^{\prime}_{ij} = \sin \theta^{\prime}_{ij}$ (i,j = 0, 1, 2). 
The v.e.v. of the exotic scalars $v_{1,2} $ determine the PQ scale and have therefore to take
a high value way above the EW scale. This implies a certain tuning of the mixed coupling $ \lambda_{h\phi_i} $.
If we consider that quartic couplings
$\lambda_{12}, \lambda_{h\phi_i}, \lambda_{\phi_i} \sim \lambda_h$ 
($i = 1, 2$) then the mixing angles between the SM
Higgs and the beyond SM Higgses become very small,
\begin{eqnarray}
\tan \theta^{\prime}_{01} \sim \frac{v}{v_1}\,,\,\,\,\,\,
\tan \theta^{\prime}_{02} \sim \frac{v}{v_2}\, ,
\end{eqnarray}
so $\theta^{\prime}_{01}, \theta^{\prime}_{02} \sim 10^{-9}$.
For this choice of parameters, the exotic Higgs fields are very heavy
$M_{h_{1,2}} \sim 10^{7}$ TeV. The other possibility would be
to consider $h_{1,2}$ in the mass range around TeV scale and mixing
with the SM Higgs $\mathcal{O}(0.1)$,
in that case, we need quartic couplings associated with the extra singlets 
scalars very small ($\sim 10^{-7}$). In this work, we consider
$h_{1,2}$ mass in between these two extreme cases {\it i.e.} in the range
TeV to a few hundreds TeV.

\section{Neutrino masses} \label{neutrino}

In the present model we can generate the light neutrino masses below the $eV$ scale 
by the Type I seesaw mechanism. 
Once the singlet scalars and the SM Higgs doublet take v.e.v.s 
(as given in Eq.\,(\ref{phih})),
the Dirac mass matrix in the basis ($\nu_{e}, \nu_{\mu}, \nu_{\tau}$)
and ($N_{2}, N_{3}$) takes the following form,
\begin{eqnarray}
m_{d} = m_{d\,fi} = 
\begin{pmatrix}
\frac{y_{e2}  v v_{1} }{2 M_{PL}} & 
\frac{(y^{R}_{e3} + i y^{I}_{e3}) v v_{1} }{2 M_{PL}}\\
\frac{y_{\mu 2} v}{\sqrt{2}} & \frac{(y^{R}_{\mu 3} + y^{I}_{\mu 3}) v}{\sqrt{2}} \\
\frac{y_{\tau 2} v}{\sqrt{2}} & \frac{(y^{R}_{\tau 3} + y^{I}_{\tau 3}) v}{\sqrt{2}}
\end{pmatrix}\,,
\label{neutrino-dirac-mass}
\end{eqnarray}
where $f = e, \mu, \tau$ and $i = 2, 3$. We can absorb the phases of
the first column in the lepton fields, but then the phases
of the second column cannot be removed. 
The elements in the first row are coming from higher dimensional operators
and are therefore suppressed by $ v_1/M_{PL} $. 
The right handed neutrino mass matrix 
in the basis ($N_2, N_3$) takes instead the following form,
\begin{eqnarray}
m_{R} =
\begin{pmatrix}
M_{22} & M^{R}_{23} + i M^{I}_{23} \\
M^{R}_{23} + i M^{I}_{23} & M_{33}
\end{pmatrix}\,,
\label{right-handed-neutrino-mass}
\end{eqnarray}
where $M_{ij} = \frac{y_{ij} v_{1} v_{2}}{2\, M_{PL}}$ ($i,j = 2, 3$)
as obtained from Eq.\,(\ref{hdo-rh-neutrino}).
The elements of the right handed neutrino mass matrix are coming from the
higher dimensional operators and give naturally RH neutrino masses at the EW
scale.
The light neutrino masses arise by the Type I seesaw mechanism as
\begin{eqnarray}
m_{\nu} &=& - m_{d} m^{-1}_{R} m^{T}_{d} 
\end{eqnarray}
The light neutrino mass matrix can be diagonalised by the PMNS
matrix consist of three mixing angles $\theta_{12}$, $\theta_{13}$,
$\theta_{23}$ and one phase $\delta$. The mixing angles and the mass
square differences of the light neutrinos have been measured
very precisely by the oscillation experiments \cite{Capozzi:2016rtj, Esteban:2020cvm}. 
The present day $3 \sigma$ bounds on the oscillation parameters 
are as follows,
\begin{itemize}
 
\item bound on mass squared differences in $3\sigma$
 range \cite{Capozzi:2016rtj, Esteban:2020cvm}
$2.37<\dfrac{\Delta m^2_{31}}
{10^{-3}}\,{\text{eV}^2} < 2.63$ and
$6.93<\dfrac{\Delta m^2_{21}}
{10^{-5}}\,{\text{eV}^2} < 7.97$,

\item $3\sigma$ bound on three mixing angles $30^{\circ}<\,\theta_{12}\,<36.51^{\circ}$,
$37.99^{\circ}<\,\theta_{23}\,<51.71^{\circ}$ and
$7.82^{\circ}<\,\theta_{13}\,<9.02^{\circ}$
\cite{Capozzi:2016rtj, Esteban:2020cvm}.
\end{itemize} 

As in our setting only two of the RH neutrinos participate in the see-saw, 
one of the light neutrinos remains massless, so that the sum of the neutrino
masses is of the order of $ 0.06 $ eV, consistent with the present
cosmological bounds \cite{Ade:2015xua, Planck:2018vyg}.

As the oscillation experiments suggest sub-eV scale neutrino mass,
we choose our parameters accordingly and the masses of the right handed 
neutrinos and the Dirac mass matrix elements can be contained in the ranges,
\begin{center} 
\begin{eqnarray}
10^{-6}\,\,{\rm GeV} \leq &m_{d\, fi}& \leq 10^{-3}\,\,{\rm GeV}\,,\nn\\
1\,\,{\rm GeV} \leq &m_{R}& \leq 100\,\,{\rm GeV}
\label{mass-range}
\end{eqnarray}
\end{center}

\begin{figure}[h!]
\centering
\includegraphics[angle=0,height=7.5cm,width=8.5cm]{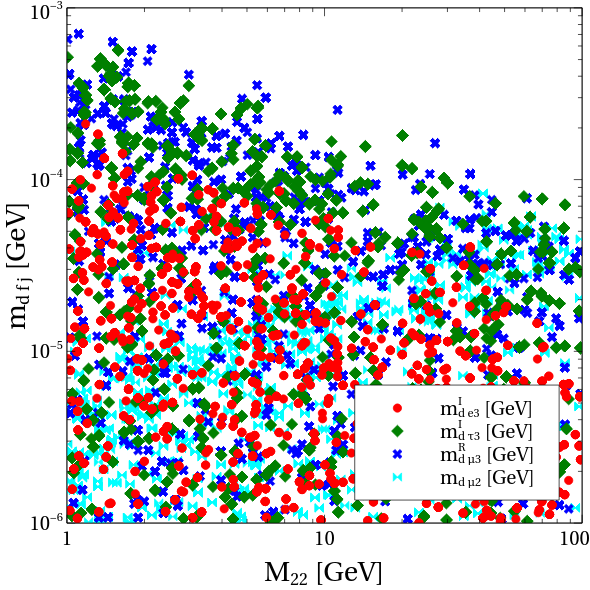}
\includegraphics[angle=0,height=7.5cm,width=8.5cm]{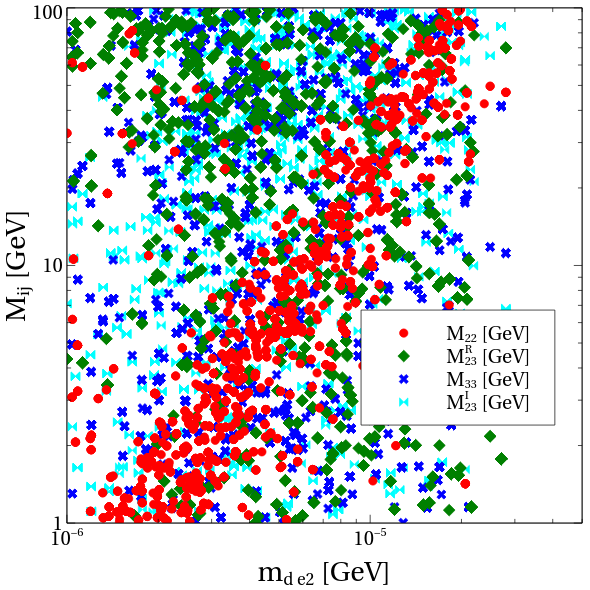}
\caption{LP shows the scatter plot among the parameters 
$M_{22} - m_{d\,\,fj}$ where $m^I_{d\,e3}$ by red circle points, 
$m^{I}_{d\,\tau3}$ by green
diamond points, $m^{R}_{d\,\mu3}$ by blue cross points and 
$m_{d\,\mu2}$ by cyan tiehorz points. 
In the RP, variation in $ M_{ij} - m^{I}_{d\,\,\mu 3} $ plane is
shown where $M_{22}$ represented by
red circle points, $M^{R}_{23}$ by green diamond points, $M_{33}$ by blue 
cross points and $M^{I}_{23}$ by cyan tiehorz points.
All the points are obtained after satisfying neutrino
oscillation data (NOD) considering the normal hierarchy of neutrino mass.} 
\label{neutrino-scatter-plot-1}
\end{figure}

In Fig.\,(\ref{neutrino-scatter-plot-1}), we show the scatter plots
among the different elements of Majorana mass matrix ($m_{R}$) 
and Dirac mass matrix ($m_d$) after fitting the neutrino oscillation data.
As can be seen, there are more than five free parameters associated with the 
neutrino mass so we can easily 
satisfy the neutrino oscillation data both for normal and inverted hierarchy. 
In this figure we show the normal hierarchy case.

As each element of the neutrino mass matrix 
$m_{\nu}$ has to be smaller than $10^{-10}$ GeV, the elements of the Dirac
mass matrix have to be small, around $10^{-3}$ GeV, while the remaining
suppression comes from the inverse Majorana mass matrix.
 In the RP, we see a nice correlation among the red points which is due
 to the bound on the mass square differences.

\begin{figure}[h!]
\centering
\includegraphics[angle=0,height=7.5cm,width=8.5cm]{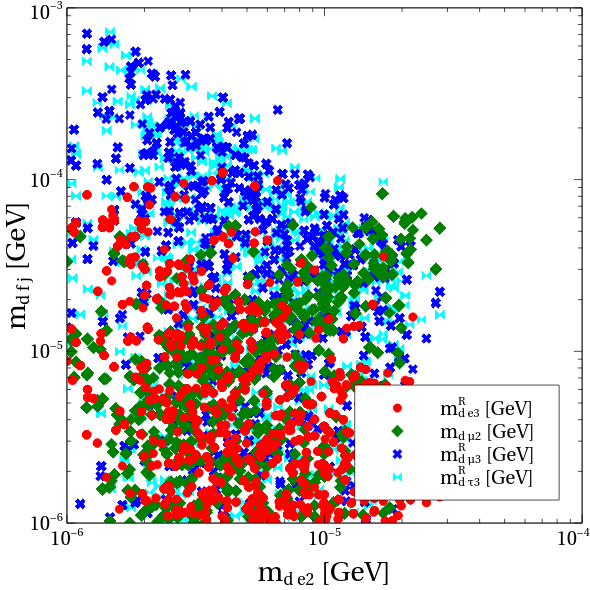}
\includegraphics[angle=0,height=7.5cm,width=8.5cm]{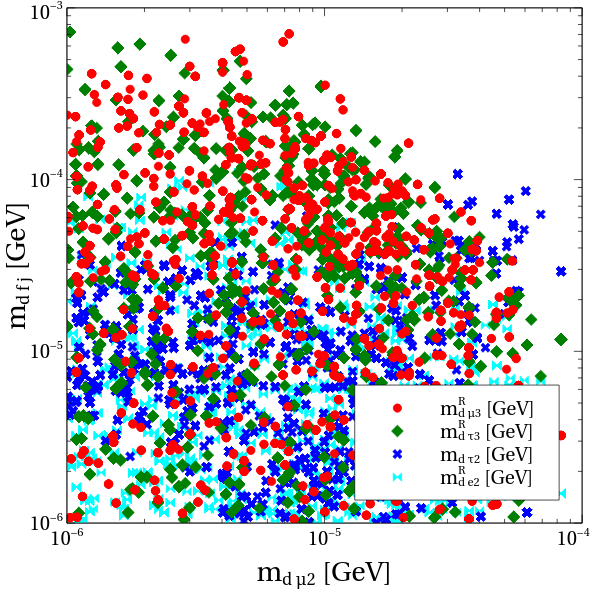}
\caption{LP shows the Scatter plot among the parameters 
$m_{d\, e2} - m_{d\, ij}$ where $m_{d\, ij}$ = $m^{R}_{d\,e3}$ 
(red circle points), $m_{d\,\mu2}$ (green diamond points), 
$m^{R}_{d\,\mu3}$ (blue cross points) and $m^{R}_{d\,\tau3}$ (cyan tiehorz points). 
In the RP,  the correlation among the $m_{d\,\mu2} - m^{R}_{d\,\mu3}$ (red circle points), $m_{d\,\mu2} - m^{R}_{d\,\tau3}$ (green diamond points),
$m_{d\,\mu2} - m_{d\,\tau2}$ (blue cross points) and 
$m_{d\,\mu2} - m_{d\,e2}$ (cyan tiehorz points) 
parameters have been shown.
All the points satisfy the neutrino oscillation data (NOD).}
\label{neutrino-scatter-plot-2}
\end{figure}
In the LP and RP of Fig.\,(\ref{neutrino-scatter-plot-2}), we show 
the variation among the different elements of the Dirac mass matrix.
Different colours correspond to different elements of the Dirac
mass matrix and have been explained in the caption of the figure.
From both the figure we can see that individual elements can reach values
up to  $10^{-3}$ GeV but only for the last two generations of light neutrinos.
The elements related to the electron neutrino are smaller.

Since the operators giving rise to the RH neutrino mass matrix are of dimension 5,
even if the scalar v.e.v.s are large, the mass matrix entries remain in the 10-100 GeV range 
and so without ad-hoc cancellations the mass eigenvalues are in the same range.

\section{Axion}
\label{axion-part}

In the model section we have given a detailed discussion of the scalar particle content, 
here we want to focus on the CP odd part of the two singlet scalars $\phi_1$ and $\phi_2$ 
and show that one combination of those plays the role of the QCD axion.

If we write the scalar fields as
\begin{eqnarray}
\phi_1 = \frac{1}{\sqrt{2}} (v_1 + H_1)  e^{i \frac{a_1}{v_1}}, \,\,\,\, \phi_2 = \frac{1}{\sqrt{2}} (v_2 + H_2)  e^{i \frac{a_2}{v_2}} 
\label{phi-axion}
\end{eqnarray} 
where $v_1$, $v_2$ are vacuum expectation values of the two fields, we can identify the
fields  $a_1$, $a_2$ with the massless Goldstone bosons.

The kinetic term for the above two scalars with the extra gauge boson $Z_{BL}$ takes the following form,
\begin{eqnarray}
\mathcal{L}^S_{kin} = -\frac{1}{4}F_{X\,\mu\nu} F_{X}^{\mu\nu} +  \sum_{i = 1, 2} (D_{\mu} \phi_i)^{\dagger} (D^{\mu} \phi_i)
\label{kin-phi1-phi2}
\end{eqnarray} 
where $F^{\mu\nu}_{X}$ is the field strength tensor for the $U(1)_{X}$ 
gauge boson $Z_{X}$ and the
covariant derivative is written as,
\begin{eqnarray}
D^i_{\mu} = \partial_{\mu} - i g_{X} Z_{X\,\mu} q_{i}
\end{eqnarray}
where $g_{X}$ and $q_i$ are the $U(1)_{X}$ gauge coupling and the charges of the corresponding fields, respectively
 $ q_1 = \alpha_L - \alpha_R, q_2 = \beta_L - \beta_R $ as in Table~\ref{u1xcharge}. The value of these charges can be
 rescaled by fixing the free parameter $ \alpha_R $ and if not otherwise specified, we choose it such to have $q_2 = 1$.
 Using the expression Eq.~(\ref{phi-axion}), Eq.\,(\ref{kin-phi1-phi2}) takes the following form,
\begin{eqnarray}
\mathcal{L}^S_{kin} = -\frac{1}{4} F_{X\,\mu\nu} F_{X}^{\mu\nu}
+ \sum_{i=1,2} \frac{1}{2} \partial_{\mu} a_i \partial^\mu a_i 
+ \frac{1}{2}  M^2_{Z_{X}} Z_{X\,\mu} Z_X^\mu 
- g_{X} Z_{X\,\mu} \partial^{\mu} \left[q_1 v_1 a_1 + q_2 v_2 a_2 \right] 
\label{kin-a1-a2}
\end{eqnarray}
where 
\begin{eqnarray}
M^2_{Z_{X}} = g^2_{X} ( q^2_1 v^2_1 + q^2_2 v^2_2 )\, .
\label{mzbl-mass}
\end{eqnarray}
Here we see that one combination of the pseudoscalars $ a_i $ mixes with the $Z_X $ gauge boson and can
be absorbed into that field as the longitudinal component of the massive gauge boson, so define the new 
pseudoscalars as,
\begin{eqnarray}
A &=&  \frac{1}{\sqrt{q^2_1 v^2_1 + q^2_2 v^2_2}} \left[q_2 v_2 a_1 - q_1 v_1 a_2 \right]\, 
= \frac{1}{\sqrt{n^2_{\chi} v^2_1 +  v^2_2}} \left[v_2 a_1 + n_{\chi} v_1 a_2 \right]\,, 
\nn \\ 
B &=& \frac{1}{\sqrt{q^2_1 v^2_1 + q^2_2 v^2_2}} \left[q_1 v_1 a_1 + q_2 v_2 a_2 \right]\,
= \frac{1}{\sqrt{n^2_{\chi} v^2_1 + v^2_2}} \left[- n_{\chi} v_1 a_1 + v_2 a_2 \right]\,.
\label{expression-A-B}
\end{eqnarray}  
where in the last expressions we have used Eq.\,(\ref{gravity-u1x}) and 
$\frac{q_1}{q_2} = \frac{\alpha_{L} - \alpha_{R}}{\beta_{L} - \beta_{R}} = - 
\frac{N_{\chi}}{N_{\psi}} = - n_\chi $. 

So the $B$ scalar is absorbed by the massive gauge boson $ Z_X $,
while the extra Goldstone boson $A$ remains massless at the scale of $U(1)_X$ breaking.

One can easily notice that the complete Lagrangian (see Eq.\,(\ref{lagrangian-model}))
has an accidental (axial for the exotic sector) $U(1) $ global symmetry with the charge assignments 
shown in Table \ref{tab2}. We call this accidental symmetry the global Peccei-Quinn (PQ) symmetry.
With a field dependent axial transformation for the exotic quarks,
\begin{center}
$\psi_{L} \rightarrow e^{ i  \frac{a_1}{2 v_1}},\,\,
\psi_{R} \rightarrow e^{- i \frac{a_1}{2 v_1}},\,\, 
\chi_{L} \rightarrow e^{i \frac{a_2}{2 v_2}},\,\,
\chi_{R} \rightarrow e^{- i \frac{a_2}{2 v_2}}$ \,,
\end{center}  
we can eliminate the pseudoscalar fields from the Yukawa couplings
and generate the axion field coupling with the gluons as follows,
\begin{eqnarray}
\mathcal{L}_{AGG} &=& 
\left(\frac{N_{\psi} a_1 }{v_{1}} + \frac{N_{\chi} a_2}{v_{2}} \right)
\frac{g^2_{s}}{32 \pi^2} G_{\mu\nu} \tilde{G^{\mu\nu}} \nn \\
&=&  N_\psi \frac{A}{F_a}\,\,\frac{g^2_{s}}{32 \pi^2} G_{\mu\nu} \tilde{G^{\mu\nu}} \;   ,
\label{L_AGG}
\end{eqnarray}
from Eq.\,(\ref{expression-A-B}), one can see that the decay constant corresponding to the CP odd particle 
$A$ is indeed $F_a =  \frac{v_1 v_2}{\sqrt{n^2_{\chi} v^2_1 + v^2_2}}$ for $ N_\psi = 1$,
and then $\theta_{a} = \frac{A}{F_a}$ varies in the range $0$ to $2\pi$. 

The PQ symmetry is broken not only by the spontaneous breaking of the $ U(1)_X$,
but also by instanton effects, such that after the QCD phase transition, the axion
has the usual periodic potential 
\begin{eqnarray}
V_{QCD} (A)= M^2_{a} F^2_a (1 - \cos^2  (\theta_a + \bar \theta) )\,,
\label{qcd-axion-potential}
\end{eqnarray} 
where $M_a $ is the axion mass and given by
\begin{eqnarray}
M_a = \frac{m_{u} m_{d}}{(m_{u} + m_{d})^2} \frac{f_{\pi} m_{\pi}}{F_a}
\end{eqnarray}
where $ m_u, m_d $ are the quark masses, while $ m_\pi  = 139.57 $ MeV is the
pion mass and $ f_\pi \sim 130 $ MeV the pion decay constant. So a v.e.v. of the axion $A$ can
cancel the $ \bar\theta $ term and we can define a field with vanishing v.e.v. as $ \bar \theta_a = \theta_a + \bar \theta $.

\subsection{Gravitational effects in the axion potential}

It is well-known that Planck scale physics does not conserve global symmetries. 
Wormholes \cite{Hawking:1987mz, Giddings:1988cx, Gilbert:1989nq} 
are one example in support of this and black holes also can 
absorb charges without effect, as predicted by the no-hair theorem \cite{no-hair}.
More explicitly, if in a scattering process a virtual or non-virtual
black hole is formed from the initial state with definite charge, 
such global charge is not conserved in the final state due to 
Hawking evaporation \cite{Hawking:1974sw}. 

In this work, we assume a gauged $U(1)_X$ symmetry as underlying
the accidental PQ symmetry $U(1)_{PQ}$.
Therefore, at the Planck scale we can expect only higher dimensional operators
which conserve the gauge symmetry but violate the global $U(1)_{PQ}$ symmetry.
Thanks to our choice of $ U(1)_X $ charges, many operators are
forbidden by the local gauge symmetry and so the lowest dimensional operator 
of this kind is of high dimension~\cite{Kamionkowski:1992mf, Fukuda:2018oco, deVries:2018mgf},
\begin{eqnarray}
V_{PL}(\phi_1,\phi_2) = \frac{g}{N_{\psi}! N_{\chi}!} \frac{\phi^{N_{\psi}}_1 \phi^{N_{\chi}}_2}{M^{N_{\psi} + N_{\chi} -4}_{PL}} + {\it h.c.}
\label{higher-dim-operator}
\end{eqnarray}
where $g = |g| e^{i \delta}$ is a complex coupling\footnote{By assuming this coupling complex, we are not
explicitly assuming that Planck scale contribution is complex. The phase can also come from the chiral rotation of the
fermions to diagonalise the mass matrix ($m_q$) and in that case this angle is $\delta \simeq arg({\rm det} [m_q])$ }.
$M_{PL} =  1.22 \times 10^{19} \,\,\mbox{GeV} $ is the Planck mass. 

Note that this operator is invariant with respect of the $U(1)_X $ local symmetry due to the anomaly
condition in eq. (\ref{su3-u1x}), but for $ N_\psi \neq N_\chi $ it breaks the PQ symmetry. 
For large number of fields $ N_\psi $ and/or $ N_\chi $, it is strongly suppressed and does not
affect the axion potential too much.
Indeed after spontaneous symmetry breaking, it gives an extra contribution to the axion potential
as follows \cite{Kamionkowski:1992mf},
\begin{eqnarray}
\mathcal{V}_{g} = (M_a^g)^2 F^2_a \left[1 - \cos(p\; \bar \theta_a + \delta) \right]
\end{eqnarray}
  where $M_a^g$ is the quantum-gravitational induced axion mass and have the following expression,
  \begin{eqnarray}
  (M_a^g)^2 = \frac{|g|}{N_{\psi}! N_{\chi}!} \frac{v_1^{N_\psi} v_2^{N_\chi} }{(\sqrt{2})^{N_{\psi} + N_{\chi}} M^{N_{\psi} + N_{\chi} - 4}_{PL} F^2_a}\,,
  \end{eqnarray}
  and  $p = - N_{\psi} +  N_{\chi} = N_\psi ( n_\chi - 1) $.
  
  So the axion potential becomes
  \begin{eqnarray}
  V(\bar{\theta}) = F^2_a  M^2_a  \left[ \left(1 - \cos \bar\theta_a \right) +
  r_g \left(1 - \cos(p\;\bar\theta_a + \delta) \right) \right]\; ,
  \label{complete-axion-potential}
\end{eqnarray}   
where $r_g = \frac{(M_a^g)^2}{M_a^2} \ll 1$.
Therefore the minimum of the axion potential (as shown in Eq.\,(\ref{qcd-axion-potential})) is shifted from 
$\bar\theta_a = 0$ due to the presence of the phase $\delta $ in the second term.
 The axion can still address the strong CP-problem if  the shift lies below the upper bound, $\Delta \theta \leq 10^{-10}$.
 By minimising the complete potential of the axion field in Eq.\,(\ref{complete-axion-potential}),
 we obtain the following value of $\bar\theta_a $ (defined it as $\Delta \theta$),
 \begin{eqnarray}
 \Delta \theta =  \frac{ r_g | p \sin \delta|}{\left[1 + p^4 r_g^2 + 2 p^2 r_g \cos \delta \right]^{1/2}} 
 \; .
 \label{delta-theta}
 \end{eqnarray}
If we assume $|p \sin\delta| \sim 1$, $p^2 r_g \ll 1 $ since for very high
v.e.v. of the singlet scalars ($v_i \sim 10^{10}$ GeV, i=1,\,2) the parameter
$r_g \ll 1$, Eq.\,(\ref{delta-theta}) takes the following form
\begin{eqnarray}
\Delta \theta \sim r_g &=&
 \frac{|g|}{N_{\psi}! N_{\chi}!\,\,(\sqrt{2})^{N_{\psi} + N_{\chi}}}
 \frac{v_1^{N_{\psi}} v_2^{N_{\chi}}}{M^{N_{\psi} + N_{\chi} - 4}_{PL} (M_a F_a)^2}
\nn\\
&=&  \frac{|g|}{N_{\psi}! N_{\chi}!\,\,(\sqrt{2})^{N_{\psi} + N_{\chi}}}
 \frac{v_1^{N_{\psi}} v_2^{N_{\chi}}}
{M^{N_{\psi} + N_{\chi} - 4}_{PL} (f_\pi m_\pi)^2} \frac{(m_u+m_d)^4}{m_u^2 m_d^2} \; .
\end{eqnarray}
So taking the values $ N_\psi = 1$ and $ N_\chi = n_\chi \gg 1 $, we have a stronger dependence
on $ v_2 $.  In the Section \ref{axion-as-dark-matter}, we show the variation of 
$\Delta \theta$ in the $v_{1} - v_{2}$ plane. 

For the special case $ v_1 = v_2  $ and $ |g| \sim 1 $ the expression gives
\begin{eqnarray}
\Delta \theta \sim  
\frac{1}{n_{\chi}!}
\left[ \frac{1+n_\chi^2}{2} \right]^{\frac{1+n_\chi}{2}}
\left[ \frac{F_a^2}{f_\pi m_\pi} \right]^2
\left[ \frac{F_a}{M_{PL}} \right]^{n_\chi-3}  \frac{(m_u+m_d)^4}{m_u^2 m_d^2}  \; .
\end{eqnarray}

Using Stirling's formula for $ n_\chi \gg 1 $, this can be reduced to
\begin{eqnarray}
\Delta \theta \sim  
\frac{e^{n_\chi}}{(\sqrt{2})^{1+n_\chi}}
\left[ 1 + \frac{1}{n_\chi^2}  \right]^{\frac{1+n_\chi}{2}}
\sqrt{\frac{n_\chi}{2\pi}} \;
\left[ \frac{F_a^2}{f_\pi m_\pi} \right]^2
\left[ \frac{F_a}{M_{PL}} \right]^{n_\chi-3}  \frac{(m_u+m_d)^4}{m_u^2 m_d^2} \; .
\end{eqnarray}
We see that we need $n_\chi $ to be sufficiently large to give a strong suppression
of the shift in the vacuum value of the $\theta $ angle, as the dominant 
behaviour is determined by the exponential of $ n_\chi (1 + \ln  \frac{F_a}{M_{PL}} ) < 0 $.
Considering $ n_\chi = 9 $, the largest number of fields compatible with asymptotic 
freedom in QCD up to the high scale, we obtain
\begin{eqnarray}
\Delta \theta \sim  29.41 \times 10^{-10} \left[ \frac{F_a}{10^{10} \mbox{GeV} } \right]^{10}
 \; .
\end{eqnarray}
so we see that we need unfortunately to introduce of the order of 10 exotic fermions
or more to satisfy the present experimental constraints without relying on fine-tuning 
or reducing the axion decay constant around or below the astrophysical bounds
  from the cooling of the 1987A Supernovae \cite{Chang:2018rso}, giving 
  $F_a \geq 2 \times 10^{8}$ GeV. We have checked numerically that
  $n_{\chi} = 8$ is ruled out from the $\theta-$parameter bound.

\begin{figure}[h!]
\centering
\includegraphics[angle=0,height=6.5cm,width=5.5cm]{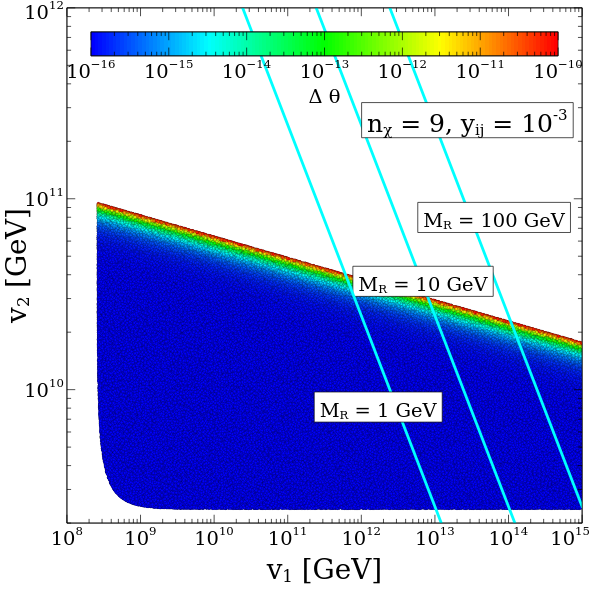}
\includegraphics[angle=0,height=6.5cm,width=5.5cm]{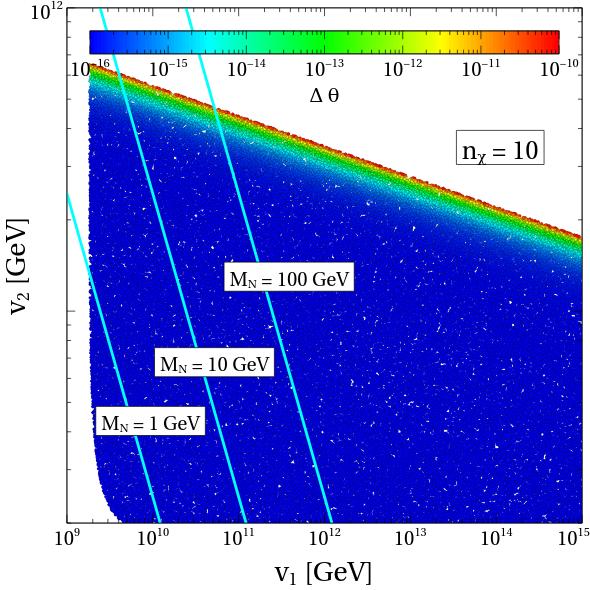}
\\
\includegraphics[angle=0,height=6.5cm,width=5.5cm]{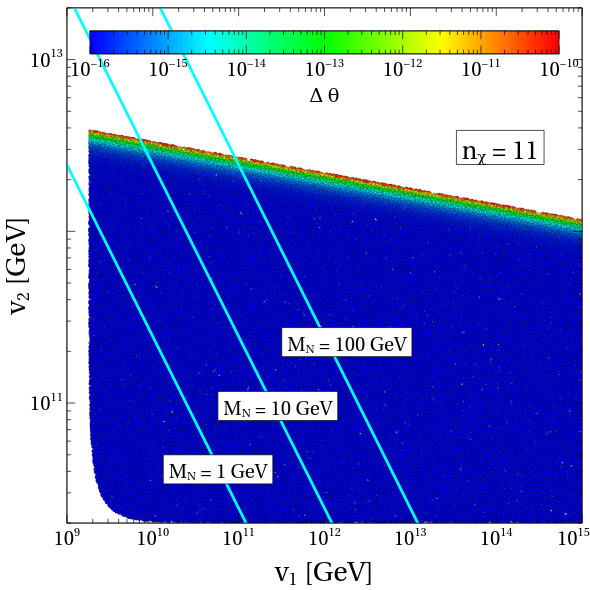}
\includegraphics[angle=0,height=6.5cm,width=5.5cm]{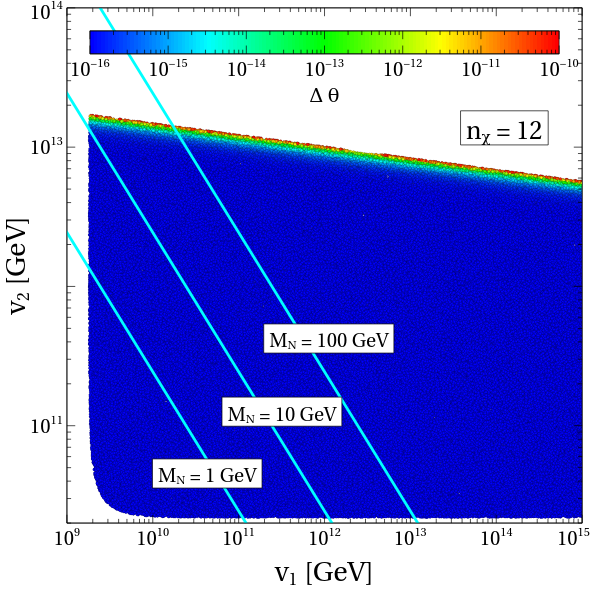}
\caption{Scatter plot in the $v_{1}-v_{2}$ plane. The value of $\Delta \theta$
parameter is shown by the color bar. Cyanide lines represent the different
mass eigenvalues of the right handed neutrino mass matrix (as defined
in Eq.\,(\ref{right-handed-neutrino-mass})) when $y_{ij} = 10^{-3}$ 
($i,j = 2, 3$). The plots from top left to right bottom 
correspond to $N_{\psi} = 1$ and $n_{\chi} = 9, 10, 11$ and $12$\,, respectively.}
\label{fig-axion-v1-v2}
\end{figure}

In Figure \ref{fig-axion-v1-v2} we show the allowed parameter space in
the $v_1$ versus $ v_2$ plane for  $ n_\chi = 9, 10, 11, 12 $ obtained numerically with the
full expression in eq.(\ref{delta-theta}).
As we can easily see from the figure, increasing the $n_{\chi}$ value, enlarges the 
allowed range of $v_2$. This is because larger $n_{\chi}$ gives a larger Planck scale 
suppression. We can also see a turnover of the blue region near 
$v_{2} \sim 2 \times 10^{10}$ which is due to the lower limit for the 
axion relic  density we have considered in the present work {\it i.e.} 
$\Omega_{a} h^{2} \sim 10^{-4}$. Moreover, the right handed neutrinos
mass also depends on the v.e.v.s $v_{1,2}$, so the cyan lines represent 
the different values of right handed neutrino mass for $y_{ij} = 10^{-3}$ 
($i,j = 2,3$). 
In the next sections we will investigate more in detail these cases
not too far from the asymptotic freedom case with respect to the 
DM density.

\subsection{Axion as Dark Matter}
\label{axion-as-dark-matter}
We consider here  a sub-eV scale axion and it can be a very good cold dark matter
candidate.
The axion contribution to the cold dark matter density from the misalignment mechanism takes the following form \cite{Turner:1985si},
\begin{eqnarray}
\Omega_{a} h^{2} \simeq 0.18\, \theta^2_{i}
\left( \frac{F_a}{10^{12}\,{\rm GeV}} \right)^{1.19}\,, 
\label{axion-dm-relic-density}
\end{eqnarray}
where the initial value $\theta_{i} \sim 1 $, as averaged over many domains if the PQ
symmetry is broken after inflation or if the initial condition is not fine-tuned for the
case of PQ symmetry breaking before inflation.

In Fig.\,(\ref{fig-axion-v1-v2}), we see that the bound from $\Delta\theta $ 
constrains the value $F_a$  below $10^{12}$ GeV, so that for most  of the parameter space, 
the axion density does not constitute the full dark matter relic density of the universe, apart
for an unnaturally large  $\theta_{i} $. 
Then we need an additional Dark Matter component or an additional axion production
mechanism. Topological defects like axion strings could contribute to DM production, 
in the case when the PQ and $ U(1)_X $ are broken after the inflationary epoch, but
the predictions are still uncertain~\cite{Gorghetto:2018myk, Buschmann:2019icd, Gorghetto:2020qws, Buschmann:2021sdq}.
Generically that additional production could move the region of axion DM to a lower
value of $ F_a$ by about one order of magnitude~\cite{Buschmann:2021sdq}.
But in our case the strings formed are also $ U(1)_X $ strings, which allows them to
also decay in other particles than axions and could reduce axion production.
We therefore consider the conservative case of main production via misalignment.

Note that our model has no domain wall problem for $ N_\psi = 1 $, even if $ N_\chi $
is large, as the smallest of the two plays the role of the overall domain wall number in
eq.~(\ref{L_AGG}) and we do not expect axion production from domain walls.

\begin{figure}[h!]
\centering
\includegraphics[angle=0,height=6.5cm,width=5.5cm]{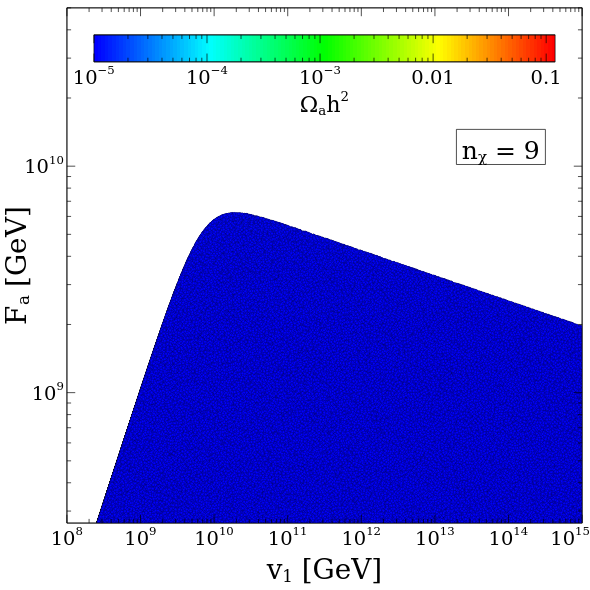}
\includegraphics[angle=0,height=6.5cm,width=5.5cm]{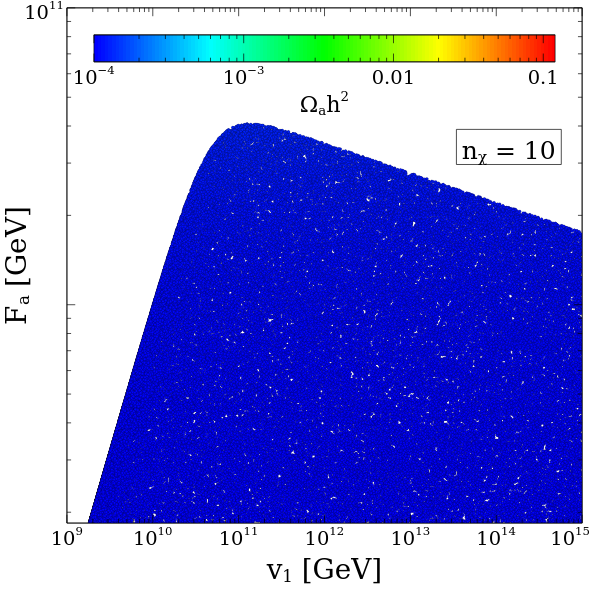}
\\
\includegraphics[angle=0,height=6.5cm,width=5.5cm]{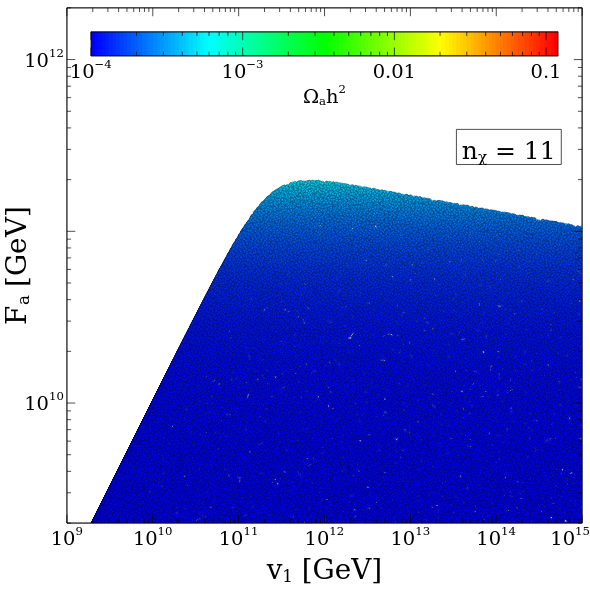}
\includegraphics[angle=0,height=6.5cm,width=5.5cm]{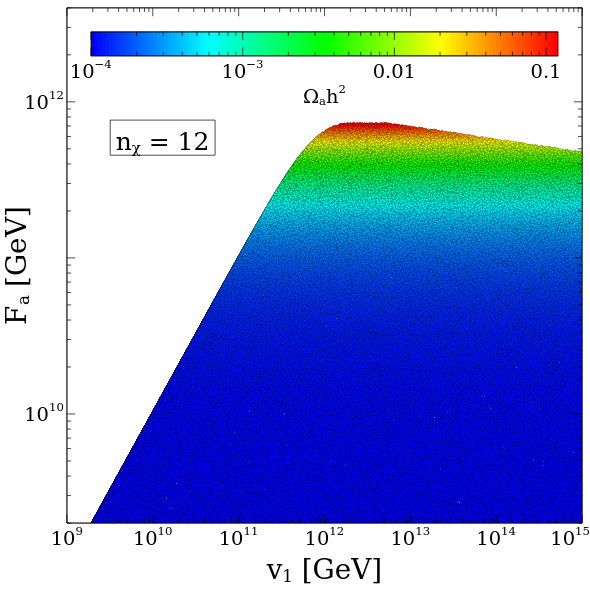}
\caption{Scatter plot in the $F_a-v_{2}$ plane. The color bar 
corresponds to axion relic density in the range $10^{-4} - 10^{-1}$. 
Moreover, the color region is allowed by the bound on $\theta$
parameter as well. The plots from top-left to bottom-right are for 
$n_{\chi} = 9, 10, 11$ and $12$\,.}
\label{fig-axion-v1-fa}
\end{figure} 

In Fig.\,(\ref{fig-axion-v1-fa}), we show the scatter plots in the $F_a - v_1$ plane 
after satisfying the bound on the  $\theta$ parameter, $\Delta \theta < 10^{-10}$
and showing in color the axion DM density via misalignment from
Eq.\,(\ref{axion-dm-relic-density}).
The four plots from top-left to bottom-right correspond to $N_{\psi} = 1$ and $n_{\chi} = 9, 10, 11$ and $12$,
respectively. For low values of $n_{\chi}$, the constraint from the neutron dipole moment
leaves open only a parameter region with a too small contribution to the DM relic density. 
For $ n_\chi = 10, 11$ the maximal value of  $F_a$ reaches
$4 \times 10^{10}, 2\times 10^{11} $ GeV respectively.

For $n_{\chi} = 12$ or larger, we can reach a maximal value $F_a = 10^{12}$ GeV, 
so the axion can provide the full DM density in the Universe.
But still in a large region of the parameter space the density is too small
and therefore we need another DM component.
In this model such component can be the third right-handed neutrino
$ N_1 $, which is stable thanks to the $\mathbb{Z}_2$ symmetry,
and will be discussed in section \ref{FIMP}

\subsection{Axion coupling with $SU(2)_{L}$ and $U(1)_{Y}$ gauge bosons}

In the present work, in order to single out one of the RH neutrinos, the first 
generation of leptons are charged under $U(1)_{PQ}$ symmetry. 
Because of these non-trivial charges and the fact that the EW couplings are
non-vectorial, we obtain non-vanishing couplings of the axion field with the 
$SU(2)_L$ gauge bosons and the $U(1)_{Y}$ gauge boson. 
As defined in Table (\ref{tab1}, \ref{tab2}),
the PQ charge associated with $L_{e}$ and $e_R$ are equal to $- 2 q_a$. 
The $W^{i}W^{i}A$ ($i = 1, 2, 3$) coupling in the present model due to the 
nontrivial contribution from the 
$SU(2)_{L} \times SU(2)_{L} \times U(1)_{PQ}$ anomaly term can be expressed as
\begin{eqnarray}
\mathcal{L}_{WWA} &=& \frac{g_2^2}{64 \pi^2}  \frac{A}{F_a} 
\tilde{W}_{\mu\nu}^i W^{\mu\nu}_i \nonumber \\
&=& \frac{g_2^2}{64 \pi^2} \frac{A}{F_a}
\tilde{W}_{\mu\nu}^i W^{\mu\nu}_i  \; .
\end{eqnarray}   
In the same way, we also obtain a non-trivial contribution from the 
$U(1)_{Y} \times U(1)_{Y} \times U(1)_{PQ}$ anomaly term which reads as,
\begin{eqnarray}
\mathcal{L}_{AYY} &=& (2 Y^2_{L e} - Y^2_{e}) \frac{g_1^2}{32 \pi^2} \frac{A}{F_a} 
\tilde{F}_{\mu\nu}^Y F^{\mu\nu\; Y} \nonumber \\
&=& - \frac{g_1^2}{64 \pi^2} \frac{A}{F_a} 
\tilde{F}_{\mu\nu}^Y F^{\mu\nu\; Y}
\end{eqnarray}
where we have used $Y_{L e} = -\frac{1}{2}$ and $Y_{e} = -1$\,. 
Combining the $W^3$ and $ B^Y$ we obtain a vanishing coupling to the photon
due to an exact cancellation. Indeed in this model the exotic fermions are not charged 
under the EM interaction, so they also do not contribute, and the leptons have
vectorial couplings with respect of the electromagnetic $U(1)$.
Nevertheless a non-vanishing coupling arises from the pion-axion mixing giving \cite{DiLuzio:2020wdo}
\begin{eqnarray}
\mathcal{L}_{A\gamma\gamma} &=& - \frac{e^2}{12 \pi^2} \left( \frac{4 m_d + m_u}{m_d+m_u} \right) 
\frac{A}{F_a} \tilde{F}_{\mu\nu} F^{\mu\nu}  \; .
\end{eqnarray}
Therefore, the axion has substantial interaction with the photons and
with the electroweak gauge bosons which can be measured at 
ongoing and future experiments~\cite{Billard:2021uyg}.
In our model the density of the axion can be lower than the total DM density, but the axion-photon 
coupling can be larger due to the constraint on $ F_a $ from $ \Delta\theta $,
so that it may be still detectable.
Unfortunately the mass range is shifted accordingly, leading to a larger mass than
those recently explored by ADMX~\cite{ADMX:2021nhd}, but those mass range could be
accessible in the future~\cite{Billard:2021uyg}.

\section{FIMP Dark Matter}\label{FIMP}

From Fig.\,(\ref{fig-axion-v1-fa}) we see that the axion cannot be the sole 
dark matter component for $ n_\chi < 12 $ and neither for the entire region in 
the $v_1 - F_a$ plane if $ n_\chi =12 $, since the EDMs bound does not allow 
large enough $ F_a$ to match the total DM density.
Fortunately, we have another DM candidate in the model since we have assigned 
$\mathbb{Z}_2$ parities to the right handed neutrinos such that $N_{1}$ 
is the lightest   $\mathbb{Z}_2$ odd particle and stable, since the Higgs fields 
do not break this global symmetry. 
In order for $N_1 $ to be a good DM candidate, it has to be produced
in the early Universe in right amount; due to its very weak interactions
a natural way for that to happen is through the FIMP mechanism \cite{Hall:2009bx}.
Indeed the only interactions of the right handed neutrino $N_1$ are the 
$U(1)_X $ gauge interaction and the non-renormalisable operator generating
its Majorana mass as
\begin{eqnarray}
\mathcal{L}_{N_1} = \frac{i}{2} \bar N_1 
\gamma^{\mu}\left(\partial_{\mu} - i\, n_{e} g_{X} Z_{X} \right) N_1 +
y_{11}  \bar N^{c}_1 N_1 \frac{\phi^{\dagger}_1 \phi_2}{M_{Pl}}+ {\it h.c.}
\end{eqnarray} 
where $n_e$ and $g_X$ are the charge of $N_1$ and gauge coupling associated 
with $U(1)_{X}$ gauge group.

We can produce the $N_1$ from the decay of scalars contained in $ \phi_{1,2} $
($h_1$, $h_2$ assuming the EW symmetry is unbroken and $ v=0 $ in the mass matrix
in eq. (\ref{neural-mass-higgs})) and of the gauge boson ($Z_{X}$). The production
of $N_1$ via the freeze-in mechanism requires feeble couplings and therefore
also the gauge coupling $ g_X $ has to be very small and the mass of the
new $Z_X$ boson from $ v_1, v_2 $ will be naturally in the TeV range.
Consequently, the extra gauge boson will not reach thermal equilibrium 
in the early Universe and will be itself produced via the FIMP mechanism,
similarly to what happens also in the model in~\cite{Biswas:2017ait}.
Once $Z_{X}$ is produced sufficiently,  it will then decay to two $N_1$ particles. 
This kind of dark matter production is similar to the superWIMP mechanism,
just with a non thermal density for the decaying particle, we can call it `superFIMP'
production.  We can determine the DM relic density once we know the
amount of $Z_{X}$ produced and solved the relevant Boltzmann equations.

For a decay process $A \rightarrow B C$, where $C$ is the DM candidate 
feebly coupling and therefore out-of-equilibrium, the relic density obtained
from the  mother particle decay in equilibrium is simply given by~\cite{Hall:2009bx}
\begin{eqnarray}
\Omega_C h^{2} = \frac{1.019 \times 10^{27} g_A}{g_s \sqrt{g_\rho}}\, \frac{M_{C} \Gamma_A}{M^2_A}\, ,
\label{hall-decay}
\end{eqnarray}
where $ g_A $ counts the internal d.o.f. of the mother particle, while $ g_s, g_\rho $ ($\sim 100$) denote the number of 
relativistic d.o.f. in the entropy and energy density of the Universe.
The DM production from the Higgs decays for $h_{i},i= 1,2$ is of this type
assuming that the scalars are in equilibrium thanks to the interaction with the SM Higgs 
and the additional quartic and cubic couplings. Indeed due to the scatterings with SM
fields, the exotic Higgs scalars are in equilibrium as long as their quartic couplings are 
larger than $ 10^{-13} $.
The RH neutrino energy density produced from these decays is therefore well-described
by the formula above taking into account that two neutrinos are produced
in each decay as
\begin{eqnarray}
\Omega_{N_1}^{FIMP}  h^{2} = \frac{2.038 \times 10^{27}}{g_s \sqrt{g_\rho}}\, 
\sum_i \frac{M_{N_1}  \Gamma_{H_i}}{M^2_{h_i}}\,.
\label{Higgs-decay}
\end{eqnarray}
Here the decay of the Higgs into neutrinos is determined by the same operator giving the neutrino mass term
and so we have for each Higgs field, assuming small mixings so that $ H_{1,2} $ are nearly equal to $ h_{1,2} $, 
\begin{eqnarray}
\Gamma_{H_i \rightarrow N_1 N_1 } =    
\frac{M_{H_i} M_{N_1}^2 }{ 16\pi \; v_i^2 } \, ,
\label{Higgs-to-N-decayrate}
\end{eqnarray}
leading to
\begin{eqnarray}
\Omega_{N_1}^{FIMP} h^{2} \sim \frac{2.038 \times 10^{27}}{g_s \sqrt{g_\rho}}\, 
\sum_i \frac{M_{N_1}^3}{16\pi M_{h_i}\; F_a^2 (n_\chi^2+1) }\,.
\label{Higgs-decay}
\end{eqnarray}
suppressed by both the Higgs mass and the scalar v.e.v. $ v_1 = v_2 = F_a \sqrt{n_\chi^2+1}$ .
We see therefore that the FIMP production from direct Higgs decay into neutrino is often
too suppressed to give a substantial DM component as long as $ M_{N_1} \leq 10^{-4} M_{h_i} $.

For the production of neutrinos from the gauge boson $ Z_X$, we have instead to consider
the coupled Boltzmann equations, as also $ Z_X $ does not equilibrate.
The production of $ Z_X $ itself takes place from the decay of the Higgs fields 
since it is lighter than the scalar fields due to the suppressed gauge coupling. 
Otherwise it can also be produced in the scatterings of the fermions of the SM, 
which are charged under $ U(1)_X $, but the scattering processes are 
negligible compared to the decays as discussed in~\cite{Biswas:2017ait}.

The Boltzmann equation to determine the phase-space distribution $ f_{Z_{X}}(p) $  
of the gauge boson $Z_{X}$ has to be solved to be able to compute its non-thermal 
average decay rate into $ N_1$ as described in \cite{Konig:2016dzg, Biswas:2017ait, Abdallah:2019svm}.
The non-thermal average of the decay width of the $ Z_X $ is indeed given by,
\begin{eqnarray}
\langle \Gamma_{Z_{X} \rightarrow N_{1} N_{1}}
\rangle_{NTH} = M_{Z_{X}} 
\Gamma_{Z_{X} \rightarrow N_{1} N_{1}}
\frac{\int \frac{f_{Z_{X}}(p)}{\sqrt{p^{2}
+M_{Z_{X}}^{2}}} d^{3} p}
{\int f_{Z_{X}}(p) d^{3} p}\,.
\label{nth-decay-zbl}
\end{eqnarray} 

In the equation for the yield, both a production and a decay term appear as they may be active 
at the same time. Using as a time variable $ z = M_{h_1}/T $, with
$ M_{h_1} $ the mass of the lighter Higgs scalar, we have the coupled equations
\begin{eqnarray}
\frac{d Y_{Z_X}}{d z} &=& 
\frac{2\, M_{PL}\, z\, \sqrt{g_{\star} (z)}}
{1.66\, M_{h_1}^{2}\, g_{s}(z)} \left[ \sum_{i=1,2} \langle 
\Gamma_{h_i \rightarrow Z_X Z_X} 
\rangle_{TH}\, (Y_{h_i} - Y_{Z_X}^2)  
-  \langle 
\Gamma_{Z_{X} \rightarrow N_{1} N_{1}} \rangle_{NTH}\, Y_{Z_{X}} \right] 
\label{be-Z}
\\
\frac{d Y_{N_1}}{d z} &=& \frac{2\, M_{PL}\, z\, \sqrt{g_{\star} (z)}}
{1.66\, M_{h_1}^{2}\, g_{s}(z)} 
\left[ \sum_{i=1,2} \langle  \Gamma_{h_{i} \rightarrow N_{1} N_{1}} \rangle Y_{h_i} 
+ \langle \Gamma_{Z_{X} \rightarrow N_{1} N_{1}} \rangle_{NTH}\, Y_{Z_{X}}  \right] 
\label{be-N3}
\end{eqnarray}
where $M_{PL} = 1.22\times 10^{19}$ GeV is the Planck mass and 
$g_{\star}(z) =
\frac{g_s(z)}{\sqrt{g_{\rho}(z)}}\left(1-\frac{1}{3}
\frac{d\,\ln g_s(z)}{d\ln z}\right)$.
Here we are neglecting the backreaction due to the $ N_1 $ density, as it is negligible
in the whole regime.

Finally, one can solve the Eq.\,(\ref{be-N3}) and determine the co-moving
 number density of $N_1$ and its relic density by the following expression,
\begin{eqnarray}
\Omega_{N_1}h^2 &=& 2.755\times 10^8
\bigg(\dfrac{M_{N_1}}{\rm GeV}\bigg)
\,Y_{N_1}(T_{\rm Now}) \,. 
\label{relic-N3}
\end{eqnarray}

\begin{figure}[h!]
\centering
\includegraphics[angle=0,height=7.5cm,width=8.5cm]{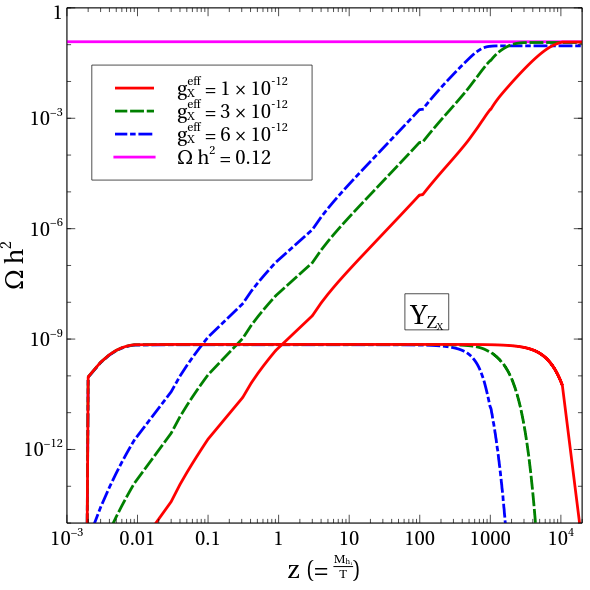}
\includegraphics[angle=0,height=7.5cm,width=8.5cm]{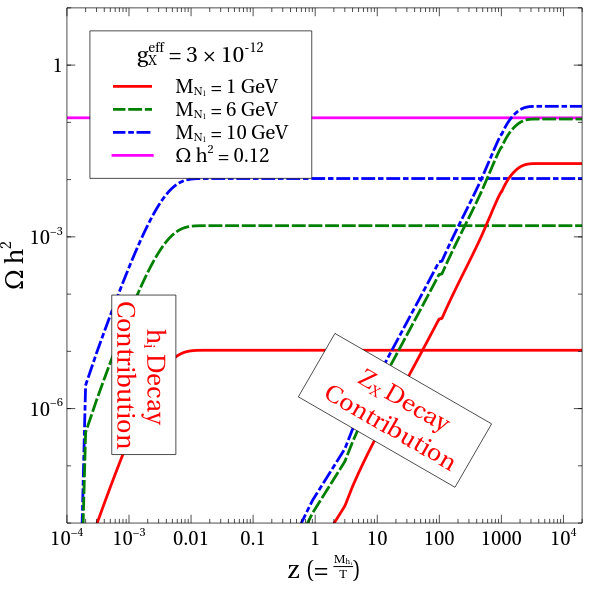}
\caption{LP: Plot of the relic density as a function of $z$ for 
GeV scale right-handed neutrino for three different values of
$g_{BL}$. The contribution comes from $Z_{BL}$ decay only and we 
have taken the model parameters
 values as follows, $M_{N_1} = 6$ GeV, $v_1 = 10^{14}$ GeV, $v_2 = 2 \times 10^{11}$ GeV, 
 $M_{h_1} = M_{h_2} = 100$ TeV.
RP: DM relic density evolution as a function of $z$, for 
$g^{eff}_{X} = 3 \times 10^{-12}$, $v_1 = 10^{14}$ GeV, 
$v_2 = 2 \times 10^{11}$ GeV,
$M_{h_1} = 100$ TeV and  $M_{h_2} = 100$ TeV. 
In both panels, the mixing angle of the neutral Higgses is
 $\theta^{\prime}_{12} = 10^{-3}$ Rad.}
\label{fig-fimp-line-plot}
\end{figure}

In Fig.\,\ref{fig-fimp-line-plot}, we show the variation of the co-moving
number density of the gauge boson $Z_{X}$ and the relic density of
the DM $N_1$. The LP shows the curves for three values of 
$g^{eff}_{X} = q_2 g_X $, which are $1\times 10^{-12}$, 
$3\times 10^{-12}$ and $6\times 10^{-12}$, respectively. 
We can notice that for all the three values of $g^{eff}_{X}$
the co-moving number density of the extra gauge boson is the same. 
This is because the decay width for the process $h_{i} \rightarrow Z_{X} Z_{X}$ decay 
is independent of the gauge coupling as in the limit of $ M_{Z_X} \ll M_{h_i} $
the main production is into the longitudinal component, equivalent to production 
of the Goldstone boson and therefore independent of the gauge coupling.

As the extra gauge boson $Z_{X}$ is produced from the decay of the Higgses $h_i$ (i = 1, 2)
in thermal equilibrium, we can determine the relic density of $Z_{X}$ on the plateau
from Eq.\,(\ref{hall-decay}). The decay rate is
\begin{eqnarray}
\Gamma_{H_i \rightarrow Z_X Z_X } =    
\frac{q^2_{i} g_X^2 M_{h_i}^3}{ 32\pi \; M_{Z_X}^2} 
= \frac{q^2_{i} M_{h_i}^3 F_a^2 }{ 32\pi \; q_2^2 v_1^2 v_2^2 } \, ,
\label{Higgs-to-ZX-decayrate}
\end{eqnarray}
giving
\begin{eqnarray}
(\Omega_{Z_X}^{FIMP} h^{2})&=&  
\frac{2.038 \times 10^{27}}{g_s \sqrt{g_\rho}}\, 
 \sum_i  \frac{M_{Z_X} q^2_{i} M_{h_i} F_a^2}{ 32 \pi q_2^2 v_1^2 v_2^2} \; .
\label{relic-Z_X}
\end{eqnarray}  
corresponding for $g^{eff}_{X} = 3\times 10^{-12}$ and the parameters as in Fig.\ref{fig-fimp-line-plot}
to $\Omega_{Z_{X}}^{FIMP} h^2 \sim 610$. 
On the other hand if we compute the $Z_{X}$ relic density using the full Boltzmann equation 
including the back-reaction as in Eq.\,(\ref{be-Z}), 
then at the plateau the co-moving number density is 
$Y_{Z_{X}} \sim 7.14632 \times 10^{-10}$,
corresponding to $ \Omega^{BE}_{Z_{X}} h^2 = 590$ using eq.~(\ref{be-N3}).
So we can conclude that the back-reaction in the $Z_X$ equation does not have a strong impact 
and that actually most of the $Z_X $ decays happen after the $ Z_X $ density has frozen-in,
at least for small $ g_X $.
Indeed, contrary to the production, the decay processes depends on the exact value of $g_{X}$, 
as $\Gamma_{Z_{X}} \propto  g^2_{X}$. So the larger $g_{X}$, the faster $Z_{X}$ decays and 
consequently the faster a $ N_1$ density is produced. As long as the $ Z_X $ production and 
decay are well separated in time, we can approximate the final density of $ N_1 $ by using
a formula analogous to the SuperWIMP mechanism, but for the FIMP $Z_X $ density as
mother particle density, i.e.
\begin{eqnarray}
(\Omega_{N_1}^{SF} h^{2})&=& 2 BR_{Z_X \rightarrow N_1 N_1} \frac{M_{N_1}}{M_{Z_{X}}} (\Omega_{Z_{X}}^{FIMP} h^{2})
= 2 BR_{Z_X \rightarrow N_1 N_1}  \frac{6 \times 611}{3000} \sim \frac{1}{10} \times \frac{3666}{3000} \sim 0.12\, ,
\end{eqnarray}  
where $ BR_{Z_X \rightarrow N_1 N_1} $ is the branching ratio of decay into two $ N_1$.
So we see that also in this case the $ N_1 $ density is independent from the exact value of the
gauge coupling $ g_X $ as long as it is small enough to avoid equilibration and separate
the production and decay epochs. Then the approximate SuperFIMP formula above
matches  the relic density obtained from the full Boltzmann equation. 

We can also have an analytic estimate of the dependence on the Higgs parameters
for the $ N_1 $ production as
\begin{eqnarray}
(\Omega_{N_1}^{SF} h^{2})&=&  
\frac{2.038 \times 10^{27}}{g_s \sqrt{g_\rho}}\, 
2 BR_{Z_X \rightarrow N_1 N_1}  \sum_i  \frac{M_{N_1} q^2_{i} M_{h_i} F_a^2}{ 32 \pi q_2^2 v_1^2 v_2^2}
\label{relic-SFpart}\\
&\sim&   
\frac{2.038 \times 10^{27}}{g_s \sqrt{g_\rho}}\, 
2 BR_{Z_X \rightarrow N_1 N_1}  \sum_i  \frac{M_{N_1} q^2_{i} M_{h_i}}{ 32 \pi q_2^2 (n_\chi^2+1)^2 F_a^2} ,
\end{eqnarray}  
determined by the v.e.v.s of the scalar fields, their masses and v.e.v.s. The last expression
is valid for the case $ v_1=v_2 = F_a \sqrt{n_\chi^2 +1} $, otherwise the largest v.e.v. dominates and 
the suppression scale becomes $n_\chi^2 v_i^2 $ instead of  $ F_a^2 (n_\chi^2 +1)^2 $.
So we see that again the neutrino energy density is suppressed by the PQ scale or the v.e.v. and $ n_\chi $,
but enhanced by a large Higgs and neutrino masses~\footnote{As also the neutrino mass and the Higgs mass
contain the v.e.v.s, we can see that at the end small couplings are needed to obtain an energy
density of neutrinos equal to that of DM, i.e. $ \lambda^{1/2} y_{11} \sim 10^{-12}$ and correspondingly 
smaller masses than $ F_a $ for the exotic Higgs fields and neutrinos respectively,}. 

The branching fraction of $ Z_X $ decay into $ N_1$'s is just determined by the charges of the SM fields,
$n, n_e $ and $ m$, assuming all other exotic states to be too heavy, i.e. we have
\begin{eqnarray}
2 BR_{Z_X \rightarrow N_1 N_1} &=&   \frac{2}{24}
\frac{(n_\chi +1)^2}{n_\chi^2 - 8 n_\chi + 28/3} \rightarrow \frac{1}{12} ,
\end{eqnarray}  
for large $ n_\chi \gg 1 $.

In the RP of Figure \,\ref{fig-fimp-line-plot}, we show the variation of DM relic density 
from the decay of the gauge boson and the Higgs fields for three different values of DM masses,
1 GeV, 6 GeV and 10 GeV
Since the v.e.v. $v_1 = 2 \times 10^{11}$ GeV, DM is produced dominantly from
the $Z_X $ decay and the Higgs contribution is negligible as visible in the figure.
For $N_1$ masses above 100 GeV, the DM becomes overabundant 
because the coupling is directly proportional to the DM mass. In considering the 
production of $N_1$ from the Higgs's decays 
we have assumed here small  mixing among the Higgses without loss of any generality 
as for larger mixing our conclusion remains same.

\subsection{\bf Scatter Plots}

In generating the scatter plots among the different parameters of the present 
model, we have varied the model parameters in the following range,
\begin{eqnarray}
10^{10} \,\,\,{\rm GeV} \leq v_{1}\,,v_{2} 
\leq 10^{15} \,\,{\rm GeV} \nn \\
10^{-14} \leq g^{eff}_{X} \leq 10^{-11} \nn \\
1\,\,{\rm GeV} \leq M_{N_1} \leq 10^{3}\,\,{\rm GeV} \nn \\
1\,\,{\rm TeV} \leq M_{h_1}, M_{h_2} \leq 10^{3}\,\,{\rm TeV}
\end{eqnarray}
In the scatterplots we have both contributions from the decay 
of the scalar particles in equilibrium ($ h_{1}, h_{2}$) 
and from the $Z_{X}$ decay not in equilibrium. 
For the latter, in order to speed up the computation, we use the
approximate expression
\begin{eqnarray}
(\Omega_{N_1}^{SF} h^{2}) =  \xi \frac{M_{N_1}}{M_{Z_{X}}}
\,(\Omega_{Z_{X}} h^{2})\,. 
\end{eqnarray}
where $\xi$ is an efficiency factor including the branching ratio, $\xi < 1$. 
We estimate the $\xi$ factor from the Fig.\,(\ref{fig-fimp-line-plot}) 
obtaining $\xi = \frac{1}{10}$, in good agreement with the branching fraction estimate above.
We compare the $ N_1$ energy density with the cosmological 3$\sigma $ band
measured by the Planck collaboration \cite{Ade:2015xua, Planck:2018vyg}.

\begin{figure}[h!]
\centering
\includegraphics[angle=0,height=7.5cm,width=8.5cm]{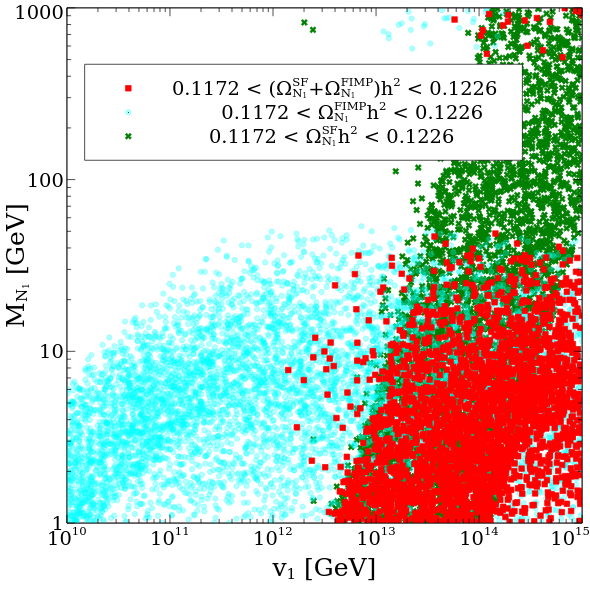}
\includegraphics[angle=0,height=7.5cm,width=8.5cm]{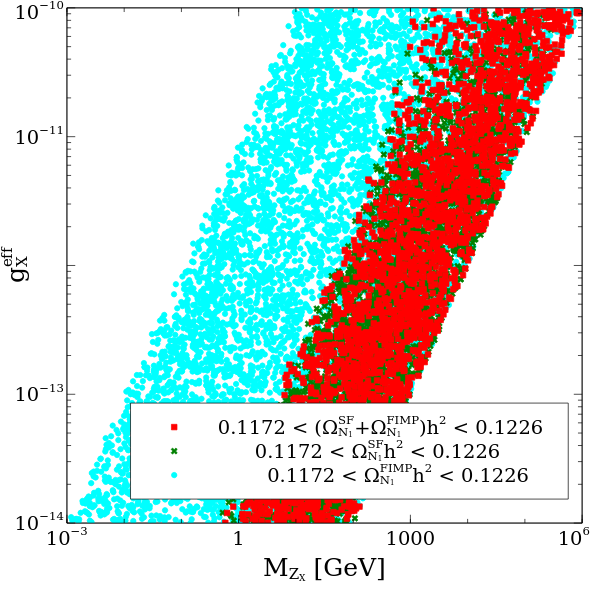}
\caption{Scatter plots in the $v_{1}-M_{N_1}$ (LP) and 
$M_{Z_{X}} - g_{X}$ (RP) planes for $ n_\chi=10$ 
after satisfying the DM relic density bound and considering production
from the decays of different parent particles. Cyan points correspond to
contribution from the scalars decays, green points correspond to 
contribution from $Z_{X}$ decay and the red points are the 
total contribution. The model parameters are fixed at
$\theta^{\prime}_{12} = 10^{-3}$ and $M_{h_{1,2}} = 100$ TeV.} 
\label{scatter-fig-1}
\end{figure}

In Fig.\,(\ref{scatter-fig-1}), we show scatter plots
in  $v_{1}-M_{N_1}$ (LP) and 
$M_{Z_{X}} - g_{X}$ (RP) planes after satisfying the bound on the $\Delta\theta$ parameter
 as well as the constraint on the DM relic density.
Cyan points represent the contribution from the scalar decays, 
green points correspond to the contribution from the $Z_{X}$ decay 
and the red points correspond to the total contribution coming from both decays.
The contribution from the scalars decay to DM is proportional to, 
$\Omega_{N_1}^{FIMP} h^{2} \propto \frac{M^{3}_{N_1}}{v_1^2} $, so we can see a linear
relation among the cyan points till $v_1 < 5 \times 10^{11}$
which is consistent with the analytical formula. 
For $v_1 > 5 \times 10^{11}$, the $ N_1 $ relic density is 
controlled by the other v.e.v. $v_2$, which is restricted by the bound from 
the $\theta$ parameter. So from that point the dependence on $ v_1 $ is
weaker and the correlation between the mass and the v.e.v. is broken.
We see that there are no allowed points for $M_{N_1} > 100$ GeV
because there the couplings
$g_{H_1 N_1 N_1} = \frac{M_{N_1}}{v_1}$ and $g_{H_2 N_1 N_1} = 
\frac{M_{N_1}}{v_2}$, become too large and the energy
density in $ N_1 $ is too large for the chosen parameters.

The contribution from $Z_{X}$ decay is proportional to the ratio of the dark matter 
mass  and the $Z_{X}$ mass as in the SuperWIMP case, so that there is
a sharp correlation among the green points. 
 The red points correspond to the total contribution which is sum of the scalar 
 and gauge boson contribution. 
 Most of the green points are not allowed because the scalars overproduce
 DM, while most of the cyan region is also ruled out because in those 
 region DM is overproduced by the decay $Z_{X}$. 
 Therefore, taking into account both contributions, only the region at large
 $ v_1 $ remains. There the value of $ F_a $ is actually determined by the
 other (smaller) v.e.v. $ v_2 $ so that the axion energy density can be
 much smaller than the $ N_1 $ energy density.
 Moreover, the DM relic density constraint can be satisfied also for higher values 
 of the DM mass, if the phase space suppression starts to play a role
 in the decay rate of the $Z_X$ boson into neutrinos reducing the BR.
 
 On the other hand,in the RP we show the points satisfying the DM energy density bound
in the  $M_{Z_{X}} - g^{eff}_{X}$ plane. The points there follow the very simple
correlation $M_{Z_{X}} \simeq g_{X} v_1$ due to the symmetry breaking. 
Here we see that for the production via Higgs decay the viable region is
larger as we vary the  $h_{1,2}$ in a broad window, but for the most part
of the parameter space the red and green point overlap, as the dominant
production comes from the $ Z_X $ decay and it is independent of the
exact gauge coupling and gauge boson mass, as given in eq. (\ref{relic-SFpart}).
Nevertheless, the gauge coupling has still to be small enough to avoid
equilibration of the gauge boson and the neutrino, $ g_X \leq 10^{-6} $,
as well as ensure $M_{h_{1,2}} > M_{Z_X}$.

 \section{Conclusion} \label{conclusion}

We have presented a new $U(1)_X $ extension of the Standard Model providing solutions and relating
three different open questions in the SM. Indeed the model contains an accidental PQ symmetry and 
an axion field solving the strong CP problem. The same scalar v.e.v.s breaking the PQ symmetry
give masses to the RH neutrinos and contributes to the light neutrino masses via the seesaw.
The RH neutrino masses are due to non-renormalisable operators, so that they are at the electroweak 
scale. Only two of the RH neutrinos take part to the seesaw mechanism, so that the lightest active 
neutrino remains massless. 
The charges of the exotic fields in the model are fixed by anomaly freedom of the new gauged
$ U(1)_X $ and these constraints also strongly reduce the number of possible non-renormalisable
operators, allowing for an accidental PQ symmetry to emerge.

In the model there are naturally two DM candidates, the axion and the only RH neutrino that does 
not couple to the active neutrinos and is stable due to the $\mathbb{Z}_2$ symmetry.
Depending on the field content of the model, the constraints on the scale $F_a $ from higher
order operators violating the PQ symmetry are such that the axion cannot provide the full
DM density and the gap has to be filled by the RH neutrino.
Different scenarios arise depending on the number of exotic quarks present in the model
encoded in the parameter $ n_\chi $ for $ N_\psi = 1 $. Such parameter fixes the order
of the gravity-induced operators that modify the $ \theta $ angle breaking the accidental
PQ symmetry.  Only the value $ n_\chi \leq 9 $ allows for the asymptotic freedom of QCD to 
survive to the high scale and in that case the EDM bounds limit the value of $ F_a$ so that
the axion cannot provide a substantial number density for Dark Matter.
Then the RH neutrino can give the total DM density for small gauge couplings $ g_X $ 
and masses in the tens of GeV with the main production from the $ Z_X $ gauge boson decay
out of equilibrium.
For values of $ n_\chi \geq 12 $ instead also mainly axion DM is possible, with a small
RH neutrino component suppressed by a small mass or large v.e.v.. Indeed the parameters
determining the two energy densities depend on different combinations of the two vacuum
expectation values and while for the axion field the key parameter is $ F_a $, determined by
the smaller v.e.v., the density of the RH neutrino is sensitive as well to the Higgs masses
and the larger v.e.v..

Since the DM density of the axion and the RH neutrino is reduced compared to the total density, 
we expect a weaker signal in direct detection experiments. 
Indeed the coupling of the axion to photons is the same as for hadronic KSVZ models with
electrically neutral exotic quarks, which is now reached by the ADMX experiment for the mass
window $ 2.6-4.2 \mu$ eV. In our case though often the axion mass is larger due to the smaller
allowed $F_a$, in the range $ 5\,\,  \mu$eV  to 28 meV and the energy density
smaller than $ \Omega_{DM} $.
The $N_1 $ instead couples to the SM fields via the $ U(1)_X $ gauge interaction with a tiny gauge
coupling. So in this scenario the new $Z_X$ gauge bosons is in the TeV range even if the
v.e.v.s of the Higgs fields is of order $ F_a $. Unfortunately the gauge coupling is
too small to allow for gauge boson production at a collider. 
On the other hand the direct detection of $N_1 $ Dark Matter via $ Z_X $ and Higgs exchange is 
independent of the small gauge coupling but suppressed by the scalar v.e.v.s and masses.
The RH neutrino sector at the EW scale can provide additional signatures of the model,
but in that case based on the standard see-saw coupling as discussed in \cite{Atre:2009rg}.
The model may be able also to include leptogenesis at the low scale via the ARS mechanism
or via resonant leptogenesis, if these two states $N_{2,3} $ are sufficiently degenerate.
Finally, the exotic scalar fields mix with the SM Higgs after EW symmetry breaking and 
they open additional Higgs decays into axions and modify generically the Higgs couplings and 
possibly as well the Higgs potential \cite{McCullough:2021odc}, but these effects are always
suppressed by the PQ scale $ F_a $ and difficult to measure.

\section{Acknowledgements}

This project has received funding from the European Union's
Horizon 2020 research and innovation programme under the Marie 
Sklodowska -Curie grant agreement No 860881-HIDDeN.
This work used the Scientific Compute
Cluster at GWDG, the joint data center of Max Planck Society for the Advancement of Science (MPG) and
University of G\"{o}ttingen.

\end{document}